\documentclass[10pt,twocolumn,twoside]{IEEEtran} 

\usepackage{color, colortbl}
\definecolor{Gray}{gray}{0.9}

\usepackage{mathrsfs}

\usepackage{stfloats}

\usepackage{cuted}

\usepackage{mathrsfs}
\usepackage{amsthm}

\newtheorem{theorem}{Theorem}

\newtheorem{lemma}{Lemma}

\newtheorem{problem}{Problem}

\usepackage{hyperref} 

\usepackage{algorithm,algorithmic}

\usepackage{amsfonts}

\usepackage{subcaption}
\usepackage{graphicx}

\usepackage[utf8]{inputenc}
\usepackage{amssymb}

\usepackage{nomencl} 
\makenomenclature
\usepackage{etoolbox}
\renewcommand{\nomgroup}[1]{%
  \item[\bfseries
  \ifstrequal{#1}{A}{Given Parameters}{%
  \ifstrequal{#1}{B}{Decision variables}{%
  \ifstrequal{#1}{C}{Other Symbols}{}}}%
]}
\usepackage{cite}

\ifCLASSINFOpdf
\else
\fi

\usepackage{amsmath}
\usepackage{algorithmic}

\usepackage{array}

\usepackage{float}

\usepackage{stfloats}

\newcommand{\pushright}[1]{\ifmeasuring@#1\else\omit\hfill$\displaystyle#1$\fi\ignorespaces}
\newcommand{\pushleft}[1]{\ifmeasuring@#1\else\omit$\displaystyle#1$\hfill\fi\ignorespaces}
\makeatother
\newif\ifmargincomments 
\margincommentstrue

\ifmargincomments

\else

\fi

\begin{document}
%




\title{Cooperative Operation of the Fleet Operator and Incentive-aware Customers in an On-demand Delivery System: A Bi-level Approach}

%
%
%





\author{Canqi~Yao\thanks{This paper is partially supported by Shenzhen Key Laboratory of Control Theory and Intelligent Systems, under grant No. ZDSYS20220330161800001; and partially supported by the Science, Technology and Innovation Com- mission of Shenzhen Municipality under Grant 20200925174707002. \emph{(Corresponding author: Shibo Chen.)}\\ \indent C. Yao is with the School of Mechatronics Engineering, Harbin Institute of Technology (HIT), Harbin, 150000, China, and is also with the Shenzhen Key Laboratory of Biomimetic Robotics and Intelligent Systems, Department of Mechanical and Energy Engineering, and the Guangdong Provincial Key Laboratory of Human-Augmentation and Rehabilitation Robotics in Universities, Southern University of Science and Technology (SUSTech), Shenzhen 518055, China  (e-mail: vulcanyao@gmail.com).}, Shibo Chen\thanks{S. Chen and Z. Yang are with the Shenzhen Key Laboratory of Biomimetic Robotics and Intelligent Systems, School of System Design and Intelligent Manufacturing, and the Guangdong Provincial Key Laboratory of Human-Augmentation and Rehabilitation Robotics in Universities, Southern University of Science and Technology (SUSTech), Shenzhen 518055, China (e-mails: shibochen.ustc@gmail.com, yangzy3@sustech.edu.cn).}, \textit{Member, IEEE,} Zaiyue Yang, \textit{Senior Member, IEEE}}

\markboth{IEEE Internet of Things Journal}%
{Yao \MakeLowercase{\textit{et al.}}: 10.1109/JIOT.2023.3324047 }
%



\maketitle

\begin{abstract}
In this paper, we study the cooperative operation problem between the fleet operator and incentive-aware customers in an on-demand delivery system. Specifically, the fleet operator offers discounts on transportation costs in exchange of customers’ delivery time flexibility. In order to capture the interaction between the fleet operator and customers, a novel bi-level optimization framework is proposed. By exploiting the strong duality, and the Karush–Kuhn–Tucker (KKT) optimality condition of customer optimization problems, we can reformulate the bi-level optimization problem as a mixed integer nonlinear programming (MINLP) problem. Considering the inherent difficulties of MINLP, a computationally efficient algorithm, which combines the merits of Lagrangian dual decomposition and Benders decomposition, is devised to solve the resulting MINLP problem in a distributed manner. Finally, extensive numerical experiments demonstrate that the proposed cooperation scheme can decrease the delivery fees for the customers, and reduce the operation cost of the fleet operator at the same time, thus leading to a win-win situation for both sides.

\end{abstract}


%
\IEEEpeerreviewmaketitle

\makenomenclature
\nomenclature[A]{$d_{ij}$}{Travel distance from node $i$ to $j$ where $i,j\in\mathcal{V}$}
\nomenclature[A]{$T_{ij}$}{Travel time from node $i$ to $j$ where $i,j\in\mathcal{V}$}
\nomenclature[A]{$\mathscr{M}_{i}$}{ Revenue from serving requests when $i\in\mathcal{R}$, set as negative value.}
\nomenclature[A]{$c_{v}$}{Vehicles' usage cost }
\nomenclature[A]{$\tau_i$}{Pickup or delivery time of transportation requests at node $i,i\in\mathcal{R}$ given by customers}
\nomenclature[A]{$\bar{\delta}_{i}$}{The maximum time flexibility provided by customer $i,i\in\mathcal{R}$}

\nomenclature[A]{$\mathcal{K}$}{The set of vehicles}
\nomenclature[A]{$\mathcal{R}$}{The set of transportation request
 nodes}
\nomenclature[A]{$M$}{A constant number used in the big-M reformulation method with large value}
\nomenclature[A]{$\Gamma$}{The value of time of the travel time}

\nomenclature[B]{$x_{ij}^k$}{Binary decision variable indicating whether $k\in\mathcal{K}$ traverses $(i,j)\in\mathcal{E}$ or not}
\nomenclature[B]{$t_i$}{Continuous decision variable indicating the arrival time at node $i,i\in\mathcal{V}$}
\nomenclature[B]{$\delta_{i}$}{The time flexibility provided by customer $i,i\in\mathcal{R}$}
\nomenclature[B]{$u_{i}$}{The dual variable of upper bound constraint $\delta_i\leq \bar{\delta_i}, \forall i\in\mathcal{R} $}
\nomenclature[B]{$\sigma_{i}$}{The dual variable of upper bound constraint  $0 \leq \delta_i, \forall i\in\mathcal{R} $}
\nomenclature[B]{$\omega_{j}^1$}{The auxiliary binary decision variable for complementarity condition  $0\leq u_j \perp (\Bar{\delta}_j-\delta_j) \geq 0,j\in\mathcal{R}$}
\nomenclature[B]{$\omega_{j}^2$}{The auxiliary binary decision variable for complementarity condition   $0\leq \sigma_j \perp \delta_j \geq 0,j\in\mathcal{R}$}
\nomenclature[B]{$\eta_{j}$}{The auxiliary continuous decision variable for  the linearization of bilinear terms in objective function $j\in\mathcal{R}$}
\nomenclature[B]{$q_{j}$}{The discount price variable of vehicles' operator for each customer $i, j\in\mathcal{R}$}


\nomenclature[C]{$G(\mathcal{V},\mathcal{E})$}{Directed graph, $\mathcal{V}=\left\{v_1,v_n \right\}\cup\mathcal{R}$ includes the start depot $v_1$, end depot $v_n$ and transportation request nodes $\mathcal{R}$, and $\mathcal{E}$ denotes the set of paths with $i,j\in\mathcal{V}$ denoting a path from vertex $i$ to $j$.}

\nomenclature[C]{$\mathcal{I}(\delta_j)$}{The inconvenience function of of customer $j$.}



\section{Introduction}


\IEEEPARstart{N}{owadays} on-demand delivery has attracted increasing attentions \cite{perboli2021new}. Consumers want their products, not just delivered to the place they prefer but also delivered at the time of their choosing. The requirement of punctuality in this novel logistic paradigm introduces great challenges to the transportation service operator, incurring considerable difficulty in solving the vehicle routing problem of its transportation fleet\cite{reyes2018complexity}. In practice, different customers hold diverse attitude towards the punctuality, i.e., some customers have high demands of fast delivery while others are more flexible about the time of delivery. Therefore, the transportation fleet operator can exploit the difference of customer flexibility in delivery time to design a more cost-effective delivery strategy. It can be achieved by demand management techniques, where the fleet operator provides discounts to customers in exchange of delivery time flexibility. This cooperation between fleet operator and incentive-aware customers can not only decrease the operation cost of the operator, but also bring financial benefits to the customers with flexibility.



\par The demand management strategy starts to be applied in the logistics industry during the last few years\cite{archetti2015multi,estrada2019biased,li2012pricing,yildiz2020pricing,YangSCE16,he2023promoting,ulmer2020dynamic,strauss2021dynamic,keskin2023dynamic}. In \cite{archetti2015multi}, a compensation mechanism is introduced in the multi-period vehicle routing problem with due dates, where customer with due dates exceeding the planning period may be postponed at a cost. Based on the previous work, Estrada-Moreno et al. \cite{estrada2019biased} introduce the possibility to offer price discounts to gain service time flexibility.  However, contrary to our research, the aforementioned papers assume that the price discounts are fixed by the operator, which cannot capture the individual preference of each customer. 
To study the single-item and uncapacitated lot-sizing problem, Li et al.\cite{li2012pricing} find out the value of offering price discounts in increasing delivery flexibility and reducing logistics cost. By extending the strategy of demand management to a variant of the multi-period vehicle routing problem, where a service provider offers a discount to customers in exchange for delivery flexibility, Yildiz et al. use an exact dynamic programming algorithm to obtain the optimal results showcasing that the cost saving up to $30 \%$ can be achieved~\cite{yildiz2020pricing}. \textcolor{black}{ Considering that the dynamic pricing scheme is able to substantially increase both revenue and the number of customers for the fleet~\cite{he2023promoting}, Ulmer et al.~\cite{ulmer2020dynamic} present an anticipatory pricing and routing policy method that incentivizes customers to select delivery deadline options efficiently for the fleet to fulfill. To further exploit the benefits of the flexible delivery time, Strauss et al.~\cite{strauss2021dynamic}  introduce flexible delivery time slots, defined as any combination of such regular time windows, and the reduced delivery charge for the customer providing flexible delivery time slots. Keskin  et al.~\cite{keskin2023dynamic} introduce a new demand management technique in dynamic vehicle routing problems, i.e., touting, in which customers that have not yet been served can be actively encouraged to order a service sooner. Besides,~\cite{keskin2023dynamic} also proposes several strategies to determine the most relevant customers to tout.}

\par However, these aforementioned papers do not consider the interaction and negotiation between the fleet operator and customers. That is, the reaction of customers is not incorporated into the vehicle routing problem solved by the fleet operator. Considering that the fleet operator and incentive-aware customers need to work cooperatively to exploit the time flexibility, neglecting the customers' reaction is not appropriate in practice. To this end, the research gap is bridged by our paper.

\par To tackle the difficulties in the vehicle routing problem solved by the fleet operator, numerous papers have proposed a set of algorithms \cite{yildiz2019provably,estrada2019biased,dayarian2016adaptive,luo2015service,rahimi2015fleet,larrain2019exact,neves2020multi,dayarian2015branch,athanasopoulos2013efficient,mohamed2019benders}. The provided solution algorithms can be roughly classified as two categories, \textit{a)} the heuristics approaches\cite{estrada2019biased,dayarian2016adaptive,rahimi2015fleet} such as search based algorithm\cite{dayarian2016adaptive,estrada2019biased,luo2015service}, and learning based heuristics method\cite{nazari2018reinforcement,lu2019learning}; \textit{b)} the exact methods such as branch and cut method\cite{larrain2019exact,neves2020multi}, branch and price method\cite{dayarian2015branch,athanasopoulos2013efficient}, column and row generation method\cite{yildiz2019provably}, and Benders decomposition method\cite{mohamed2019benders}. In \cite{estrada2019biased}, a metaheuristic approach is proposed to find low-cost solution, which includes both the transportation costs and the cost of the price discounts offered. To determine a distribution plan to visit a set of customers, Larrain et al. \cite{larrain2019exact} embed a new family of valid inequalities within the framework of branch and bound to improve its performance.  However, there are still a few limitations associated with these solution approaches. For instance, the heuristics based algorithms obtain the near optimal solution with a good performance on computation time, but without the theoretical guarantee on the solution quality. \textcolor{black}{ Regarding exact algorithms such as the Benders decomposition method, they can indeed yield optimal solutions but typically demand a substantial number of iterations. This can result in a significant computational burden. Hence, it is possible to substantially reduce the computation time of the Benders decomposition method by decreasing the number of iterations~\cite{bodur2017strengthened}. This motivation leads to the approach proposed in the subsequent sections. }

\par In this paper, we focus on the cooperative vehicle routing problem of a transportation fleet operator and incentive-aware customers.
To achieve a better routing schedule, the fleet operator provides a transportation fee discount to the incentive-aware customers in exchange for delivery time flexibility. Note that with the adoption of demand management strategy into the routing problem, the following two-fold benefits emerge, \text{(1)} customers will receive the delivery cost savings with the discount offered; \text{(2)} when the customers provide flexible delivery time windows to the operator, more flexibility in deciding the optimal routing schedule can be obtained, leading to a reduced operation cost for the fleet operator. 

\par The main contributions of this paper are presented as follows
\begin{itemize}
    \item To the best of authors' knowledge , a novel bi-level vehicle routing model characterizing the cooperation between fleet operator and incentive-aware customers is proposed for the first time in the field of vehicle routing problem. 
    \item Due to the NP hardness of bi-level vehicle routing problem, exact reformulation techniques based on the KKT optimality condition are proposed, which is used to transform the bi-level vehicle routing problem as a single-level mixed integer nonlinear programming problem (MINLP). 
    \item In order to further reduce the computation complexity, these nonlinear terms in MINLP can be exactly linearized by exploiting the property of strong duality. Then the MINLP is equivalently converted into a simpler mixed integer programming (MIP) problem.
    \item To tackle with the inherent complexities of the MIP and protect the privacy of customers, we devise a novel decomposition method named as Benders dual decomposition algorithm \textcolor{black}{integrating the Lagrangian dual decomposition method with the generalized Benders decomposition method. In comparison to the generalized Benders decomposition method, the Benders dual decomposition method requires fewer iterations, leading to significant savings in computation time, which is theoretically proved and substantiated through simulation results. Furthermore, with the strengthened valid cuts, the proposed approach can provide the optimal solution and also achieve the distributed implementation.} 
    
    
\end{itemize}

\par The rest of this paper is summarized as follows. In Section II, We elaborate the system models, and the mathematical model of bi-level vehicle routing problem considering the flexible time windows. In order to tackle the NP hardness brought by the bi-level framework and computation difficulties from nonlinear terms in the objective function, we propose a set of accurate reformulation approaches to convert the primal bi-level mathematical model into the single-level MIP based mathematical model in Section III. In addition, a computationally efficient decomposition based algorithm, which combines the complementary advantages of Benders decomposition and Lagrangian dual decomposition method, is proposed in Section IV. We conduct simulation experiments in Section V to prove the validity of proposed model and algorithm. Finally, in Section VI we concludes this paper. In this paper, customers and requests are used interchangeably.

\section{ Bi-level Model of fleet Operator and Customers }

\begin{figure}[h]
\centering
\includegraphics[width=.9\linewidth]{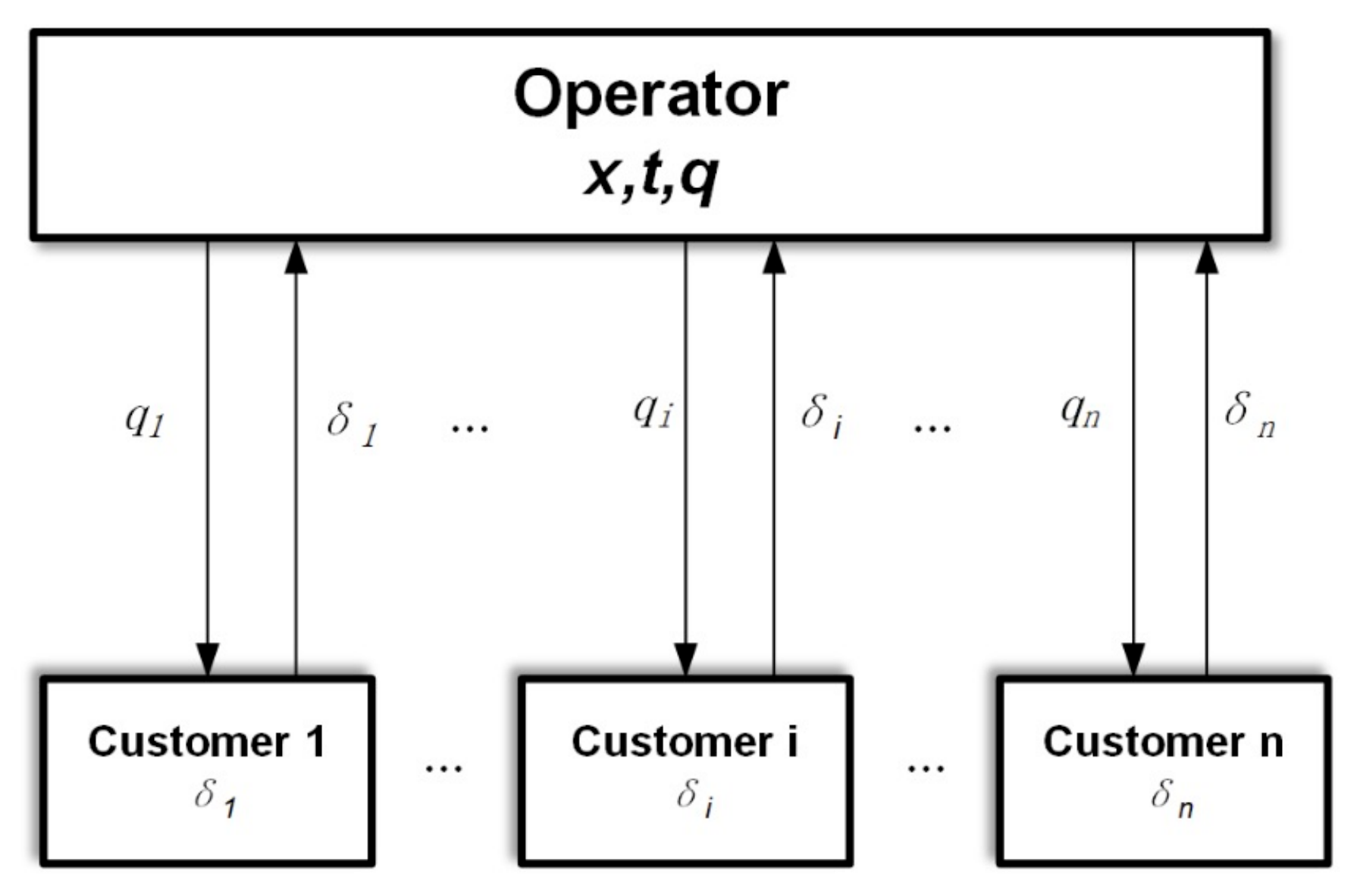}
\caption{The interaction between a fleet operator and massive transportation requests.}
\label{bi-level}
\end{figure}

In this paper, a novel business model is considered where the transportation fleet operator and incentive-aware customers work cooperatively to achieve cost-effective goods delivery. To be specific, the fleet operator 
In practice, among the scenario where a fleet operator serves massive customers, customers set the expected service start time (delivery or pickup time) individually before the fleet operator assigns specific vehicle to serve them.  As for customers who are insensitive to the service start time, to obtain the delivery (pickup) time flexibility for making better routing schedule, the fleet operator is prone to provide the service fee discount. Meanwhile customers who provide time flexibility for fleet operator will receive the delivery fee reduction.  
In this paper, we propose a novel pricing mechanism to guide the behavior of fleet operator and customers. 
Unlike the similar papers considering the strategy of demand management \cite{yildiz2020pricing,estrada2019biased,YangSCE16}, a bi-level optimization model is proposed to characterize the interaction between a company with massive customers and a fleet operator, which is shown in Fig.~\ref{bi-level}. In the following subsection, the details concerning lower-level customers model (follower problem) and upper-level operator model (leader problem) are introduced firstly, then we present the bi-level model.

\subsection{Mathematical Model of Customers}
As the followers, to reduce the delivery fee, customers are willing to provide their time flexibility $\delta_i$ in exchange for a discount price $q_i$ set by the leader (fleet operator). In addition, $\delta_i$ could also cause inconvenience for customers due to the dispersed delivery time window and delayed delivery time. Then, we propose a convex function $\mathcal{I}(\cdot)$ to quantify the inconvenience. Specifically, with the discount price $q_i$ given by the leader problem, the follower problem for each customer is shown as follows

\begin{problem}[Consumer problem --- Follower problem]\label{customer_model}
\begin{equation}
\begin{aligned}
\min_{\delta_j}\qquad   &  \mathcal{I}(\delta_j)-q_j \delta_j,    \\\text{s.t.}\quad &  0 \leq \delta_j \leq \Bar{\delta}_j:\sigma_j,u_j
\end{aligned}
\end{equation}
\end{problem}

\indent Considering that the time sensitivity varies over different customers, a lower bound ($0$) and upper bound ($\bar{\delta_j}$) constraint is also included in Problem~(\ref{customer_model}). Besides, $\sigma_j$ and $u_j$ are the corresponding dual variables for these two constraints. Since the customer model Problem~\eqref{customer_model} satisfies the Slater's condition for the convexity of~$ \mathcal{I}(\delta_j)-q_j \delta_j$ and the convex box constraints~$0 \leq \delta_j \leq \Bar{\delta}_j$ in which  $\bar{\delta_j}$ is positive, the strong duality holds for Problem~\ref{customer_model}~\cite{boyd2014convex}.

\subsection{Mathematical Model of Fleet Operator}
\par \textcolor{black}{We propose to use a directed graph to characterize the relation between depots and customers. The transportation network is modeled as a directed graph $G(\mathcal{V},\mathcal{E})$, where $\mathcal{V}=\{v_1,v_n\}\cup\mathcal{R}$ denotes the start depot $v_1$, the end depot $v_n$, and the set of customer nodes $\mathcal{R}$, as well as $\mathcal{E}$ stands for the set of paths with $i,j\in\mathcal{V}$ denoting a path from node $i$ to node $j$. We denote the start depot and end depot by $v_1$ and $v_n$, respectively. Let $T_{ij}$ be the travel time of paths between nodes $i,j\in \mathcal{V}$. Besides, the binary variable $x_{ij}^k$  denotes whether vehicle $k$ is assigned to traverse path $ij$. Besides, $c_i$ stands for a unified cost vector consisting of both the negative delivery fee $-\mathscr{M}_i$ and the vehicle usage fee $c_{v}$:
		$c_i=\left\{
		\begin{array}{ll}
			{-\mathscr{M}_i,} & {\text { if } i\in\mathcal{R}} \\ 
			{c_{v},} & {\text { if } i=v_1}
		\end{array}
		\right.
		$.}

\par 
The fleet operator, as a leader, aims to minimize its operation cost comprising of: \textit{(i)} monetary value of travel time $\sum_{k\in \mathcal{K}}\sum_{i\in\mathcal{V}} \sum_{j\in\mathcal{V}} \Gamma T_{ij}x_{ij}^k$ with $\Gamma$ denoting the monetary value of time, \textit{(ii)} the unified delivery cost defined above $\sum_{k\in \mathcal{K}}\sum_{i\in\mathcal{V}} \sum_{j\in\mathcal{V}} c_i x_{ij}^k$, and \textit{(iii)} accrued discount cost $  \sum_{j\in \mathcal{R}} q_j \delta_j \left(\sum_{i\in \mathcal{V}}\sum_{k\in\mathcal{K}} x^k_{ij}\right)  $ provided to the customers.

\begin{problem}[Transportation service operator model --- Leader problem]\label{P:operator}
 \begin{equation}\notag\label{obj}
\begin{aligned}
\min\limits_{x_{i j}^k, t_j, q_j  } \sum_{k\in \mathcal{K}}\sum_{i\in\mathcal{V}} \sum_{j\in\mathcal{V}}  & ( \Gamma T_{ij}+c_{i}) x^k_{i j} +  \sum_{j\in \mathcal{R}} q_j \delta_j \left(\sum_{i\in \mathcal{V}}\sum_{k\in\mathcal{K}} x^k_{ij}\right)  
  \end{aligned}
\end{equation}

  \begin{equation}\label{Cons_flow}
\begin{aligned}
\text{s.t.\quad}  \sum_{j\in\mathcal{V}} & x^k_{i j}-\sum_{j\in\mathcal{V}} x^k_{j i}=b_{i}, \quad \forall  i \in \mathcal{V}, k\in\mathcal{K};\\ &b_{v_1}=1, b_{v_n}=-1, b_{i|i\neq v_1,v_n}=0
\end{aligned}
\end{equation}
\begin{equation}\label{Cons_visit}
    \sum_{k\in\mathcal{K}}\sum_{j\in\mathcal{V}} x^k_{ij} \leq 1,
    \quad\forall i\in\mathcal{R}
\end{equation}

\begin{equation}\label{Cons_time}
	\begin{aligned}
		t_{j}  \geq  &T_{ij}+t_{i} -M(1-x^k_{ij}), \\ & \forall i\in \mathcal{V}\setminus  v_n, j \in \mathcal{V}\setminus  v_1,k\in\mathcal{K}
	\end{aligned}
\end{equation}

\begin{equation}\label{Cons_time-TW}
	\begin{aligned}
	\tau_{j} \leq	t_{j} \leq  \tau_{j} + \delta_{j} ,  \forall  j \in \mathcal{V}\setminus  v_1
	\end{aligned}
\end{equation}

\end{problem}
\noindent where in~\eqref{Cons_flow}, $b_{i|i\neq v_1,v_n}=0$ means that when $i\neq v_1,v_n$, $b_i=0$. 

\par In order to satisfy the physical requirements of transportation system, vehicle flow constraints, pickup (delivery) time constraints, as well as flexible time window constraints are formulated as (\ref{Cons_flow})-(\ref{Cons_time-TW}) in above mathematical formulation. Constraint (\ref{Cons_flow}) showcases the flow conservation constraint of vehicles. Specifically, a vehicle entering in the customer node has to exit out at the same customer node, and vehicles return to the end depot after starting at the start depot. In constraint (\ref{Cons_visit}), each customer is served at most once by the vehicle.  Time characteristics of customer are specified in (\ref{Cons_time}), which states that the arrival time at customer nodes should not be later than the prescribed delivery time of customers. \textcolor{black}{Besides, in \eqref{Cons_time}, a constant $M$ with a large value has been introduced to linearize the original nonlinear constraint~$t_{j}  \geq  (T_{ij}+t_{i})x^k_{ij}, \forall i\in \mathcal{V}\setminus  v_n, j \in \mathcal{V}\setminus  v_1,k\in\mathcal{K}$. } In addition, the flexible time window constraint is characterized by (\ref{Cons_time-TW}).

\textcolor{black}{In the settings of Problem~\ref{P:operator}, we assume that there are only one start and end depots in the proposed model. In order to handle the real case of the multiple depots, the following multi-depot transformation rule is devised as follows: }

 \textcolor{black}{Multi-depot transformation rule: If there are $m$ start depots and $m$ end depots in the considered fleet operation problem, we initially introduce a virtual start and end depots which are connected with the actual start and end depots of all vehicles. Subsequently, the travel time and route selection variables between the virtual depots and the actual depots are set as zero and one, respectively. Finally, all vehicles commence from the virtual start depot and conclude at the virtual end depot.}

\begin{figure}[h]
\centering
\includegraphics[width=.9\linewidth]{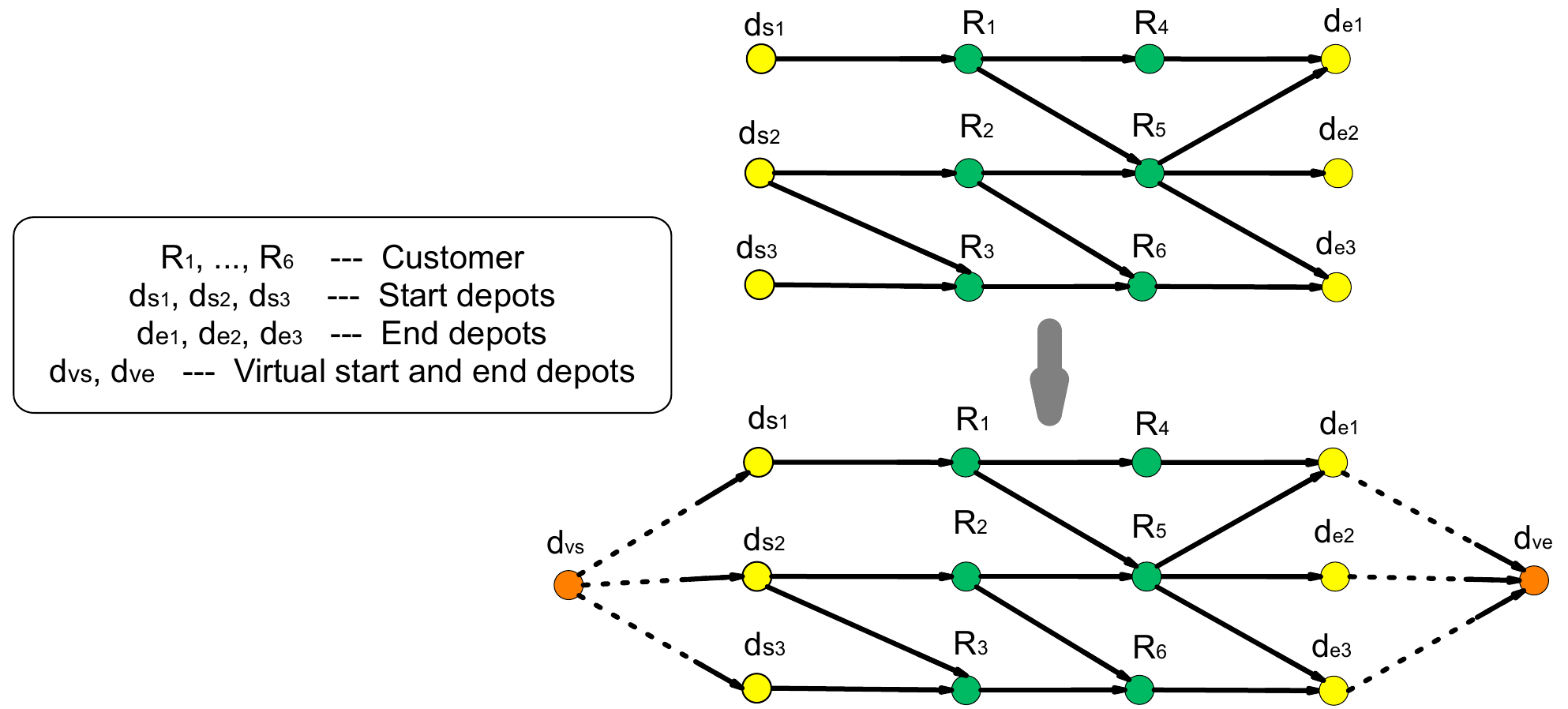}
\caption{\textcolor{black}{The fleet operation problem with three start and end depots, and the implementation of multi-depot transformation rule.}}
\label{multi-depotRule}
\end{figure}

\textcolor{black}{In Fig.~\ref{multi-depotRule},  we consider such a fleet operation problem in which there are three start and end depots, and the implementation of  multi-depot transformation rule is also illustrated. With the proposed multi-depot transformation rule, the fleet operation problem with multiple depots can be converted into the single-depot case and addressed by the proposed model.}

\subsection{Bi-level Model of Operator and Customers}
\par Considering the interaction between the fleet operator and customers, a bi-level optimization framework is employed to characterize the cooperative vehicle routing problem with flexible time window. With the mathematical formulation of customers model and fleet operator model in above subsections, the bi-level vehicle routing problem (BVRP) is formulated as follows, 
\begin{problem}[BVRP]\label{BVRP}
 \begin{equation}\notag
\begin{aligned}
\min\limits_{x_{i j}^k, t_j, q_j  }  \sum_{k\in \mathcal{K}}\sum_{i\in\mathcal{V}} \sum_{j\in\mathcal{V}}  &( \Gamma T_{ij}+c_{i})x^k_{i j} + \sum_{j\in \mathcal{R}} q_j \delta^*_j \sum_{i\in \mathcal{V}}\sum_{k\in \mathcal{K}} x^k_{ij}  
  \end{aligned}
\end{equation}
\begin{equation}\notag
\begin{aligned}
s.t.\quad\delta^*_j \in arg\min_{\delta_j}  \Big\{  &  \mathcal{I}(\delta_j)-q_j \delta_j ,   0\leq\delta_j \leq \Bar{\delta}_j \Big\}, \forall j \in \mathcal{R} 
\end{aligned}
\end{equation}
$$\textit{and}\quad \text{(\ref{Cons_flow})-(\ref{Cons_time-TW})}$$
\end{problem}
 where the customers problems are constraints to the upper level operator problem, such that, the only members that are considered feasible must be both lower level optimal and satisfy upper level constraints.

\textcolor{black}{\subsection{ Discussions}
Here, the background of proposed mathematical model, and the difficulties of solving such a model are clarified as follows:\\ \indent 
Note that the strict delivery time window may rule out the possibility of finding a good routing strategy. For example, even two customers that are geographically close may need to be served by two vehicles due to stringent delivery time constraints. Against this backdrop, the price discount is proposed as an incentive to encourage customers to provide time flexibility to the fleet operator, assisting the operator to lower its vehicle routing costs. To capture the cooperation that benefits both the fleet operator and customers, a mathematical model based on the bilevel optimization is developed. \\ }
\indent  However, even the simplest bi-level optimization problem is challenging to solve~\cite{dempe2015bilevel}. Moreover, in addition to the difficulties brought by the framework of bi-level model, there is another complicating term $\sum_{j\in\mathcal{R}}q_j \delta^*_j $, which consists of the product of two continuous decision variables, in the objective function of BVRP. In the following section, a set of reformulation techniques are carefully devised in order to equivalently transform the original bi-level problem as a single-level MIP based optimization problem.

\section{Mathematical reformulation of bi-level optimization model}
In this section, 
the BVRP is reformulated as a single-level MIP with \textit{a)} the customer problem transformation which is achieved by equivalently replacing the customer problem with its KKT optimality condition, and \textit{b)} the exact linearization approach of the nonlinear objective function. 
\subsection{Mathematical Reformulation of Customer Problem}
The bi-level optimization problem, even for the convex case, has been shown to be NP-hard\cite{dempe2015bilevel,liu2021optimal}. Hence, to tackle the NP-hardness of the bi-level model,  we propose to represent the follower problem, a convex optimization problem, with its KKT optimality condition. Firstly, a set of non-negative dual variables ($u_j,\sigma_j$) are introduced for the constraints of customers model. As a result, the KKT condition of customers problem is derived and shown in (\ref{Optimality Con}).

\begin{subequations}\label{Optimality Con}
\begin{align}
& \nabla\mathcal{I}(\delta_j) -q_j -\sigma_j +u_j =0,\forall j\in\mathcal{R} \label{Optimality Con stationary_delta}\\
& 0\leq u_j \perp (\Bar{\delta}_j-\delta_j) \geq 0, \forall j\in\mathcal{R}\label{Optimality Con complement1}\\
 &0\leq \sigma_j \perp \delta_j \geq 0,\forall j\in\mathcal{R}\label{Optimality Con complement2}
 \end{align}
\end{subequations}

\noindent where the expression $"a \perp b "$  means at most one of \textit{a} and \textit{b} can take a strictly nonzero value with the other value being 0.  The stationarity conditions are specified in  Eq.(\ref{Optimality Con stationary_delta}). These primal feasibility condition, dual feasibility condition, and complementary condition are showcased in constraints (\ref{Optimality Con complement1}), and (\ref{Optimality Con complement2}). Then original BVRP is reformulated as a single-level complicated vehicle routing problem (SCVRP) Problem~\ref{Nonlinear_MIP}.

\begin{problem}[SCVRP]\label{Nonlinear_MIP}
\begin{equation}\notag
\begin{aligned}
\min\limits_{x_{i j}^k, t_j, q_j  }  \sum_{k\in \mathcal{K}}\sum_{i\in\mathcal{V}} \sum_{j\in\mathcal{V}}  &( \Gamma T_{ij}+c_{i})x^k_{i j}  + \sum_{j\in \mathcal{R}} q_j \delta_j \sum_{i\in \mathcal{V}}\sum_{k\in \mathcal{K}} x^k_{ij}  
  \end{aligned}
\end{equation}
$$\textit{s.t.}\qquad \eqref{Cons_flow}-\eqref{Optimality Con} $$

\end{problem}

\par Note that there are still some hard-to-solve terms both in the objective function and constraints of Problem~\ref{Nonlinear_MIP}. Thus, we propose a set of exact linearization approaches to handle the difficulties from these nonlinear terms of Problem~\ref{Nonlinear_MIP} in the later subsection.

\subsection{Exact Linearization Approach for Nonlinear Terms}
In this subsection, in order to decrease the computation time of Problem~\ref{Nonlinear_MIP}, which is an MINLP problem, we propose a set of equivalent linearization methods, \textit{a)} linearizing the nonlinear complementary constraint by introducing additional binary variables, and \textit{b)} exploiting the power of strong duality to accurately linearize the nonlinear term $q_j\delta_j$.

\subsubsection{Linearization of complementary optimization constraint}
Due to the nonlinear nature of complementary constraints (\ref{Optimality Con complement1}), and (\ref{Optimality Con complement2})  which render Problem~\ref{Nonlinear_MIP} hard to solve, we linearize these nonlinear terms by introducing $2|\mathcal{R}|$ additional auxiliary binary variables $\omega^1_j,\omega^{2}_j$ and a sufficiently large constant $M$, yielding the following disjunctive constraints\cite{fortuny1981representation}

\begin{subequations}\label{Optimality Con linearized}
\begin{align}
&
\left.
\begin{aligned}
&0\leq\Bar{\delta}_j-\delta_j \leq M \omega^{1}_j\\
&0\leq u_j \leq M (1- \omega^{1}_j)
\end{aligned}
\right\},  \forall j\in\mathcal{R}
\\&
\left.
\begin{aligned}
	&0\leq \delta_j \leq M \omega^{2}_j\\
	&0\leq \sigma_j \leq M (1- \omega^{2}_j)
\end{aligned}
\right\},  \forall j\in\mathcal{R}
\end{align}
\end{subequations}

\subsubsection{Linearization of nonlinear terms in objective function} There is still a nonlinear term $q_j\delta_j$ in the objective function of Problem~\ref{Nonlinear_MIP}.  For the customer model Problem~\ref{customer_model}, in which $\mathcal{I}(\delta_j)$ is a convex function, the Lagrangian function can be obtained as shown below:
\begin{equation}\notag
\begin{aligned}
L(\delta_j,u_j,\sigma_j) = &\mathcal{I}_j(\delta_j)-q_j \delta_j - \sigma_j \delta_j  +u_j (\delta_j-\Bar{\delta}_j)\\& = \mathcal{I}_j(\delta_j)  +(u_j -q_j - \sigma_j  )\delta_j- u_j \Bar{\delta}_j
\end{aligned}
\end{equation}

The dual function can be obtained as follows\footnote{For convenience, we denote $\phi(\delta_j) =   \mathcal{I}_j(\delta_j)  +(u_j -q_j - \sigma_j  )\delta_j $},
\begin{equation}\notag
\begin{aligned}
g(u_j,\sigma_j) = &\inf_{\delta_j}  \mathcal{I}_j(\delta_j)  +(u_j -q_j - \sigma_j  )\delta_j- u_j \Bar{\delta}_j \\
& = - u_j \Bar{\delta}_j+ \inf_{\delta_j}  \phi(\delta_j)\\
& = - u_j \Bar{\delta}_j+ \phi^*(\delta^*_j)
\end{aligned}
\end{equation}

According to the strong duality of convex optimization problem, the objective value of lower level problem~\ref{customer_model} and the objective value of its dual problem are equal at the optimal solution. Then, the complicating nonlinear term $q_j\delta_j$ can be exactly linearized as follows,
\begin{equation}\notag
\begin{aligned}
 \mathcal{I}_j(\delta_j)-q_j \delta_j  &= - u_j \Bar{\delta}_j+ \phi^*(\delta^*_j)\\
     q_j \delta_j&= \mathcal{I}_j(\delta_j)+ u_j \Bar{\delta}_j- \phi^*(\delta^*_j)
\end{aligned}
\end{equation}

Then, the nonlinear terms $q_j\delta_j$ in the objective function of Problem~\ref{Nonlinear_MIP} can be exactly linearized by replacing $q_j\delta_j$ with $\mathcal{I}_j(\delta_j)+ u_j \Bar{\delta}_j- \phi^*(\delta^*_j)$. As a result, to handle the resulting bilinear terms $ (\mathcal{I}_j(\delta_j)+ u_j \Bar{\delta}_j- \phi^*(\delta^*_j))x^k_{ij} $ in the objective function, Big-M method is employed. Specifically, a continuous auxiliary $\eta_{j}$ and addition constraint (\ref{obj_linearized}) are introduced.
\begin{equation}\label{obj_linearized}
\eta_{j}\geq\mathcal{I}_j(\delta_j)+ u_j \Bar{\delta}_j- \phi^*(\delta^*_j)  -M(1-\sum_{k\in\mathcal{K}}\sum_{i\in\mathcal{J}} x^k_{ij}),\forall j\in\mathcal{R} 
\end{equation}

 Following above linearization techniques, Problem~\ref{Nonlinear_MIP} can be equivalently reformulated as the single-level tractable vehicle routing problem (STVRP) Problem~\ref{Relaxed_MIP} which essentially is an MIP problem.

 \begin{problem}[STVRP]\label{Relaxed_MIP}
 \begin{equation}\notag
\begin{aligned}
&\min\limits_{\mathbb{X} } \sum_{k\in \mathcal{K}}\sum_{i\in\mathcal{V}} \sum_{j\in\mathcal{V}}   ( \Gamma T_{ij}+ c_{i}) x^k_{i j} + \sum_{j\in\mathcal{R}} \eta_{j} 
  \end{aligned}
\end{equation}
$$\textit{s.t.}\qquad \eqref{Cons_flow}-\eqref{Cons_time-TW},\text{ and }  \eqref{Optimality Con stationary_delta}, \eqref{Optimality Con linearized},\eqref{obj_linearized} $$
 \end{problem}

\noindent where we denote $\mathbb{X}=\{ x^k_{i j}, t_j, q_j,\delta_j, u_j,\sigma_j, \omega^1_j, \omega^{2}_j,\eta_{j}   \}$ for simplicity.  
\textcolor{black}{However, since the transportation service operator and customers are independent entities, customers may not be willing to provide private information to the operator, thus hindering the centralized implementation of the solution method.} 
Hence, to protect the privacy of both sides, we propose a decomposition based solution algorithm to solve Problem~\ref{Relaxed_MIP} in the following section. Finally, to help the reader understand the underlying idea behind the proposed equivalent 
reformulation approach, a flowchart is shown as follows: 

 \begin{figure}[h]
\centering
 \includegraphics[width=.9\linewidth]{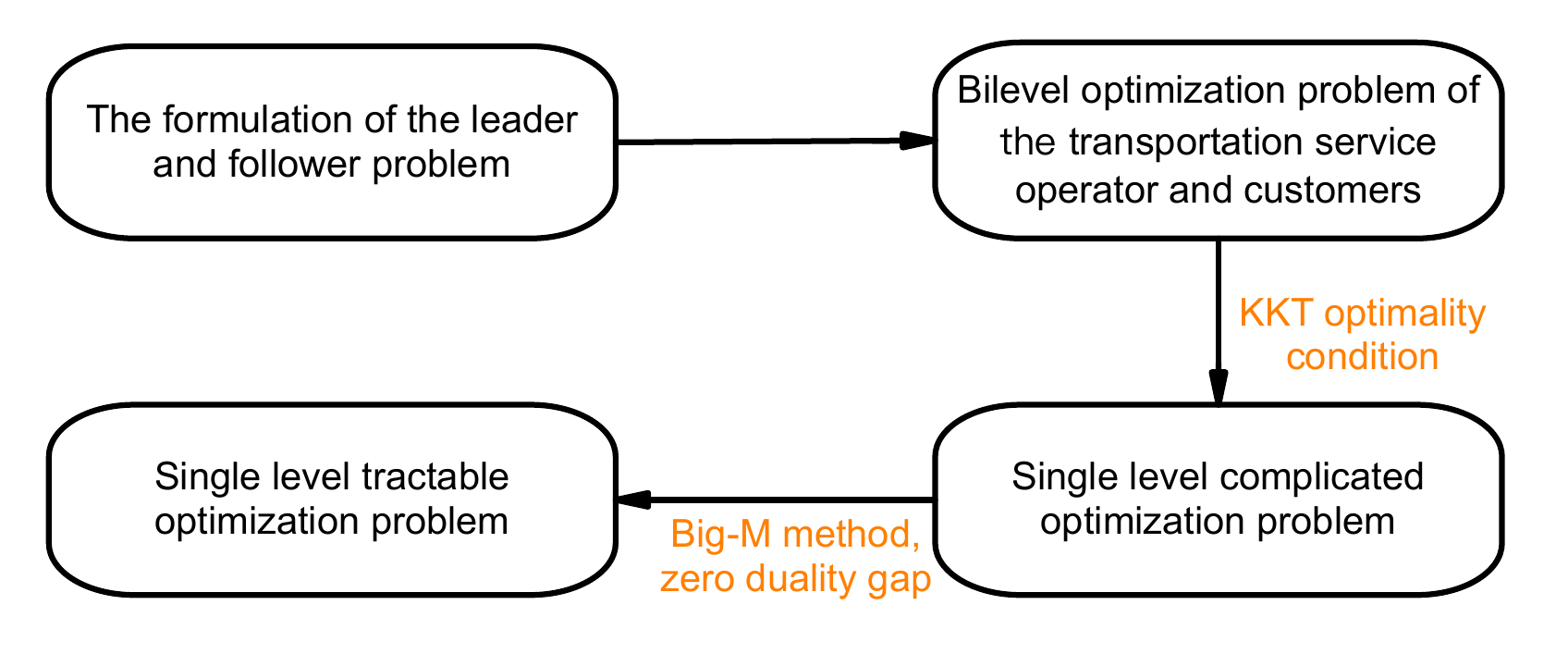}
\caption{The equivalent reformulation approach}
\label{Model:refo}
\end{figure}


\section{Decomposition based approach}
\textcolor{black}{In the aforementioned section, the bilevel optimization problem of transportation service operator and customers has been exactly converted into a single level tractable mixed integer programming problem which can be directly solved by commercial solvers.  However, it is difficult to gather the parameters of the single level tractable formulation from the transportation service operator and customers simultaneously, since they are independent entities with their own interests. Thus, to achieve a global solution, we devise a decomposition based algorithm named as Benders dual decomposition (BDD) method, which solves the resultant
Problem~\ref{Relaxed_MIP} in a distributed manner.} Note that the BDD method combines the complementary merits of Lagrangian dual decomposition and Benders decomposition method\cite{bodur2017strengthened,rahmaniani2020benders,zou2019stochastic,cerisola2009stochastic}. Specifically, by exploiting the power of Lagrangian dual decomposition method, we devise a novel form of subproblem which is an MIP problem. With the newly formulated subproblem, valid Benders cuts are generated which are stronger than the classic Benders cuts derived from the generalized Benders decomposition (GBD) method . 

\subsection{Generalized Benders Decomposition Approach}
Considering that the BDD method is based on the GBD method, thus, to demonstrate the performance of proposed BDD approach, the details of GBD method are illustrated firstly as follows\cite{geoffrion1972generalized,sahinidis1991convergence,rahmaniani2020benders}
\subsubsection{Master problem}
By relaxing constraints \text{(\ref{Cons_time})-(\ref{Cons_time-TW})}, \text{and}  (\ref{Optimality Con stationary_delta}), (\ref{Optimality Con linearized}), (\ref{obj_linearized}), the resulting master problem \textbf{(MP)} which is a relaxed version of Problem~\ref{Relaxed_MIP}, provides the lower bound for Problem~\ref{Relaxed_MIP}\footnote{In later section, we denote $m\in \mathbb{M}=\{1,2\}$}.

 \begin{equation}\tag{\textbf{MP}}\label{MP}
	\begin{aligned}
		&\min\limits_{x_{ij}^{k}, \omega_j^{m}\in \mathbb{B} } \sum_{k\in \mathcal{K}}\sum_{i\in\mathcal{V}} \sum_{j\in\mathcal{V}}  \Theta + (c_{i} +\Gamma T_{ij}) x^k_{i j} 
	\end{aligned}
\end{equation}
$$\textit{s.t.}\qquad \eqref{Cons_flow}-\eqref{Cons_visit}$$
where $\Theta$ is an auxiliary variable representing the lower bound of primal subproblem \ref{PSP}.

\subsubsection{Primal Benders subproblem}
With the integer solution obtained from solving \ref{MP}, the primal subproblem can be formulated as follows with the copies of master problem decision variables $\mathcal{X}_{ij}^k$, $\Omega_{j}$, which has been used in previous research\cite{zou2019stochastic,boland2016proximity,geoffrion1972generalized,flippo1993decomposition}.

\allowdisplaybreaks

 \begin{equation}\tag{\textbf{PSP}}\label{PSP}
	\begin{aligned}
		&\min\limits_{ \mathbb{X}_{psp} } \sum_{j\in\mathcal{R}}  \eta_{j}  
	\end{aligned}
\end{equation}

\begin{subequations}\label{Cons_psp}
	\begin{align}
s.t.\quad &\tilde{x}_{ij}^{k}= \mathcal{X}_{ij}^{k},\quad\forall i,j\in \mathcal{V}, k\in\mathcal{K} : \lambda_x^{ijk} \label{Cons-psp-copyX}\\
&\tilde{\omega}_j^{m}=\Omega_j^{m}, \quad\forall j\in\mathcal{R}, m\in\mathbb{M}: \lambda_{\omega}^{jm} \label{Cons-psp-copyW} \\
&	\begin{aligned}\label{Cons-psp-flow}
 \sum_{j\in\mathcal{V}}   \mathcal{X}^k_{i j}&-\sum_{j\in\mathcal{V}}  \mathcal{X}^k_{j i}=b_{i}, \quad \forall  i \in \mathcal{V}; k\in\mathcal{K}\\ &b_{v_1}=1, b_{v_n}=-1, b_{i|i\neq v_1,v_n}=0,
\end{aligned}\\
&	\sum_{k\in\mathcal{K}}\sum_{j\in\mathcal{V}} \mathcal{X}^k_{ij} \leq 1,
\quad\forall i\in\mathcal{R}\\
&	\begin{aligned}
	t_{j}  \geq  T_{ij}+t_{i} &-M(1- \mathcal{X}^k_{ij}),  \forall i\in \mathcal{V}\setminus  v_n, \\
	&j \in \mathcal{V}\setminus  v_1, k\in\mathcal{K}
\end{aligned}\\
&	\begin{aligned}
	\tau_{j} \leq	t_{j} \leq  \tau_{j} + \delta^*_{j} ,  \forall  j \in \mathcal{V}\setminus  v_1
	\end{aligned}\\
&\nabla\mathcal{I}(\delta_j) -q_j -\sigma_j +u_j =0,\forall j\in\mathcal{R} \\
& \left.
\begin{aligned}
	&0\leq\Bar{\delta}_j-\delta_j \leq M \Omega^{1}_j\\
	&0\leq u_j \leq M (1- \Omega^{1}_j)
\end{aligned}
\right\},  \forall j\in\mathcal{R}
\\&
\left.
\begin{aligned}\label{Cosn-KKT}
	&0\leq \delta_j \leq M \Omega^{2}_j\\
	&0\leq \sigma_j \leq M (1- \Omega^{2}_j)
\end{aligned}
\right\},  \forall j\in\mathcal{R}
\\&
\begin{aligned}
 \eta_{j}\geq\mathcal{I}_j(\delta_j)&+ u_j \Bar{\delta}_j- \phi^*(\delta^*_j) \\&-M(1-\sum_{k\in\mathcal{K}}\sum_{i\in\mathcal{J}} \mathcal{X}^k_{ij}),\forall j\in\mathcal{R}\label{Cosn-biobj-linearize}
\end{aligned}
	\end{align}
\end{subequations}


\noindent where $\tilde{x}_{ij}^{k}$, $\tilde{\omega}_{j}^{m}$ are the solution of current master problem.  We use $\mathbb{X}_{psp} = \{\mathcal{X}_{ij}^k,\Omega^m_{j}\in [0,1]  \} \cup \mathbb{X}\setminus \{x_{ij}^{k}, \omega_j^{m} \} $ to denote the feasible region of \ref{PSP}.


\subsection{Benders Dual Decomposition Approach}
\par To reduce the computation time of the GBD method, by exploiting the property of Lagrangian dual decomposition, we propose a novel decomposition method, BDD algorithm, which iteratively strengthens the master problem with stronger optimality and feasibility cut.

\par Specifically, in order to strengthen the classic generalized Benders cut, two constraints (\ref{Cons-psp-copyX}) and (\ref{Cons-psp-copyW}) are relaxed into the objective function by introducing two dual multipliers  $\lambda^{ijk}_x$ and $\lambda_{\omega}^{jm}$. The Lagrangian dual problem is formulated as follows,

 \begin{equation}
	\begin{aligned}
		\max_{\lambda} \min\limits_{\mathbb{X}_{psp} } \qquad &\sum_{k\in \mathcal{K}}\sum_{i\in\mathcal{V}} \sum_{j\in\mathcal{V}}  \eta_{j}   - \sum_{i,j\in \mathcal{V}}\sum_{k\in \mathcal{K}} \lambda_x^{ijk} (\mathcal{X}_{ij}^{k}- x_{ij}^{k*})  \\
		&-\sum_{j\in \mathcal{R}}\sum_{m\in \mathbb{M}} \lambda_{\omega}^{jm} (\Omega_j^{m}-\omega_j^{m*})\\
		&\textit{s.t.}\qquad \eqref{Cons-psp-flow},\cdots, \eqref{Cosn-biobj-linearize}\\
	\end{aligned}
\end{equation}

By applying this relaxation step, integrality
requirements can be imposed on any subset
of the variables $\mathcal{X}_{ij}^k$ and $\Omega_{j}^{m}$ given they are no longer equal to $x_{ij}^{k*}$ and $\omega_j^{m*}$ respectively. In Lemma.\ref{Lemma-opt-cut}, it is shown that a valid optimality cut can be generated by optimizing an MIP based subproblem.

\begin{lemma}\label{Lemma-opt-cut}
	Given the linear relaxation solution of master problem $x_{ij}^{k*},\omega_{j}^{m*}\in\mathbb{R}$ , and dual multipliers $\lambda_{\omega}^{jm}, \lambda_x^{ijk}\in\mathbb{R}$,  solving the Lagrangian relaxation based subproblem \eqref{Problem-LDSP}, a MIP problem, provides the  optimal solution $\{ \bar{\Omega}_{j}^m, \bar{\mathcal{X}}_{ij}^k ,\bar{\eta}_{j}   \}$. 
	 \begin{equation}\tag{\textbf{LSP}}\label{Problem-LDSP}
	\begin{aligned}
	 \min\limits_{\mathbb{X}_{psp} } \qquad &\sum_{k\in \mathcal{K}}\sum_{i\in\mathcal{V}} \sum_{j\in\mathcal{V}}  \eta_{j}   - \sum_{i,j\in \mathcal{V}}\sum_{k\in \mathcal{K}} \lambda_x^{ijk} (\mathcal{X}_{ij}^{k}- x_{ij}^{k*})  \\
		&-\sum_{j\in \mathcal{R}}\sum_{m\in \mathbb{M}} \lambda_{\omega}^{jm} (\Omega_j^{m}-\omega_j^{m*})\\
		&\textit{s.t.}\qquad \eqref{Cons-psp-flow},\cdots, \eqref{Cosn-biobj-linearize},   \mathcal{X}_{ij}^k,\Omega^m_{j} \in \mathbb{B}\\
	\end{aligned}
\end{equation}

	Then, 
	\begin{equation}\label{BDD-Optimality-cut}
		\begin{aligned}
				 \Theta \geq  &\sum_{k\in \mathcal{K}}\sum_{i\in\mathcal{V}} \sum_{j\in\mathcal{V}}  \bar{\eta}_{j}   + \sum_{i,j\in \mathcal{V}}\sum_{k\in \mathcal{K}} \lambda_x^{ijk} ( x_{ij}^{k}-\bar{\mathcal{X}}_{ij}^{k}) \\&+ \sum_{j\in \mathcal{R}}\sum_{m\in \mathbb{M} } \lambda_{\omega}^{jm} (\omega_j^{m}-\bar{\Omega}_j^{m}) 
		\end{aligned}
	\end{equation}
	is a valid optimality cut for the master problem (\ref{MP}).
\end{lemma}

The proof is in Appendix \ref{Proof-opt-cut}.


Compared with the valid cut generated by \ref{PSP}, the effectiveness of proposed valid cut (\ref{BDD-Optimality-cut}) has been rigorously proven in Theorem.\ref{TightCut}.

\begin{theorem}\label{TightCut}
Given the dual multipliers  from solving \ref{PSP},  the optimality cut (\ref{BDD-Optimality-cut}) is parallel to generalized Benders optimality cut and at least $\Xi \geq 0$ units tighter, where

	\begin{equation*}
	\begin{aligned}
		\Xi =& 	 \min\limits_{\mathbb{X}_{psp}  }  \Big\{ \sum_{k\in \mathcal{K}}\sum_{i\in\mathcal{V}} \sum_{j\in\mathcal{V}}  \eta_{j}   - \sum_{i,j\in \mathcal{V}}\sum_{k\in \mathcal{K}} \hat{\lambda}_x^{ijk} \mathcal{X}_{ij}^{k} \\ &-\sum_{j\in \mathcal{R}}\sum_{m\in \mathbb{M}} \hat{\lambda}_{\omega}^{jm} \Omega_j^{m}      \textit{:}  \eqref{Cons-psp-flow} - \eqref{Cosn-biobj-linearize} ),\mathcal{X}_{ij}^k,\Omega^m_{j} \in \mathbb{B}  \Big\}  \\ & -  \min\limits_{\mathbb{X}_{psp}  }  \Big\{ \sum_{k\in \mathcal{K}}\sum_{i\in\mathcal{V}}  \sum_{j\in\mathcal{V}}  \eta_{j}   - \sum_{i,j\in \mathcal{V}}\sum_{k\in \mathcal{K}} \hat{\lambda}_x^{ijk} \mathcal{X}_{ij}^{k}\\ & -\sum_{j\in \mathcal{R}}\sum_{m\in \mathbb{M}} \hat{\lambda}_{\omega}^{jm} \Omega_j^{m}      \textit{:} \eqref{Cons-psp-flow} - \eqref{Cosn-biobj-linearize}   \Big\} 
	\end{aligned}
\end{equation*}
\end{theorem}

The proof is in Appendix \ref{Proof-Th1}.

As for the feasibility valid cut, the same rule applies.
\begin{lemma}\label{Lemma-feasi-cut}
	Given the \textbf{infeasible} linear relaxation solution of master problem $x_{ij}^{k*},\omega_{j}^{m*}\notin\mathbb{X}_{psp}$ , and dual multipliers $\lambda_{\omega}^{jm}, \lambda_x^{ijk}\in\mathbb{R}$,  solving the \ref{FSP}, which is a MIP problem and shown in Appendix \ref{Feasibility-SP},  provides the  optimal solution $\{ \bar{\Omega}_{j}^m, \bar{\mathcal{X}}_{ij}^k   \}$.

	Then, 
	\begin{equation}\label{BDD-feasibility-cut}
		\begin{aligned}
			0 \geq  &\sum_{k\in \mathcal{K}}\sum_{i\in\mathcal{V}} \sum_{j\in\mathcal{V}}  S_{1ijk}  +  \sum_{k\in \mathcal{K}}\sum_{i\in\mathcal{V}} \sum_{r\in\{15,16\}} S_{rik}  \\&+\sum_{j\in\mathcal{V}} \sum_{r\in\{2,\cdots,14,17\}} S_{rj}+\sum_{i,j\in \mathcal{V}}\sum_{k\in \mathcal{K}} \lambda_x^{ijk} ( x_{ij}^{k}-\bar{\mathcal{X}}_{ij}^{k}) \\
			&+ \sum_{j\in \mathcal{R}}\sum_{m\in \mathbb{M} } \lambda_{\omega}^{jm} (\omega_j^{m}-\bar{\Omega}_j^{m}) 
		\end{aligned}
	\end{equation}
	is a valid feasibility cut for the master problem.
\end{lemma}

With these stronger Benders cuts \eqref{BDD-Optimality-cut}, and \eqref{BDD-feasibility-cut}, compared with the GBD method, the performance of proposed BDD method is theoretically enhanced, which will be validated in later simulation results.  \textcolor{black}{The BDD method is able to achieve the global optimal solution, which is shown in Theorem~\ref{Th_global_optimality}.}

\begin{theorem}\label{Th_global_optimality}
\textcolor{black}{
    With the strengthened Benders cuts \eqref{BDD-Optimality-cut}, and \eqref{BDD-feasibility-cut}, the proposed BDD method can still obtain the global optimal solution .  }
\end{theorem}
\begin{proof}
    \textcolor{black}{As demonstrated in Lemma~\ref{Lemma-opt-cut} and Lemma~\ref{Lemma-feasi-cut}, the proposed optimality and feasibility cuts are valid cuts for the master problem which has been proven in Appendix~\ref{Proof-opt-cut}.  This implies that the proposed cuts do not eliminate any feasible integer solutions~\cite{wolsey1999integer}. Consequently, our approach can attain the optimal solution, similar to the generalized Benders decomposition method~\cite{sahinidis1991convergence}. 
}
\end{proof}

In the subsequent subsection, we provide clarification on some implementation details.

\subsection{Implementation Details of Benders Dual Decomposition Approach}
\textcolor{black}{
Given the theoretical results clarified in Theorem.\ref{TightCut}, the solution of linear programming (LP) relaxation \ref{MP} can derive stronger Benders cut than the solution of \ref{MP}.}

\subsubsection{Fractional solution}   \textcolor{black}{At the early stage of the proposed iterative algorithm}, to quickly derive valid cuts, we ﬁrst solve the LP relaxation of the \ref{MP} with classical cuts, which is a strategy that was originally devised by McDaniel and Devine \cite{mcdaniel1977modified,rahmaniani2020benders};

\subsubsection{Stronger cut generation}The strengthened Benders cuts (\ref{BDD-Optimality-cut}), (\ref{BDD-feasibility-cut}) are generated in the following steps, \textit{1)} the value of dual multipliers $\lambda$ is obtained by solving the \ref{PSP}; \textit{2)} with the dual multipliers $\lambda$, stronger Benders cuts are generated by solving \ref{Problem-LDSP}.

\begin{figure*}[h]
\centering
 \includegraphics[width=.9\linewidth]{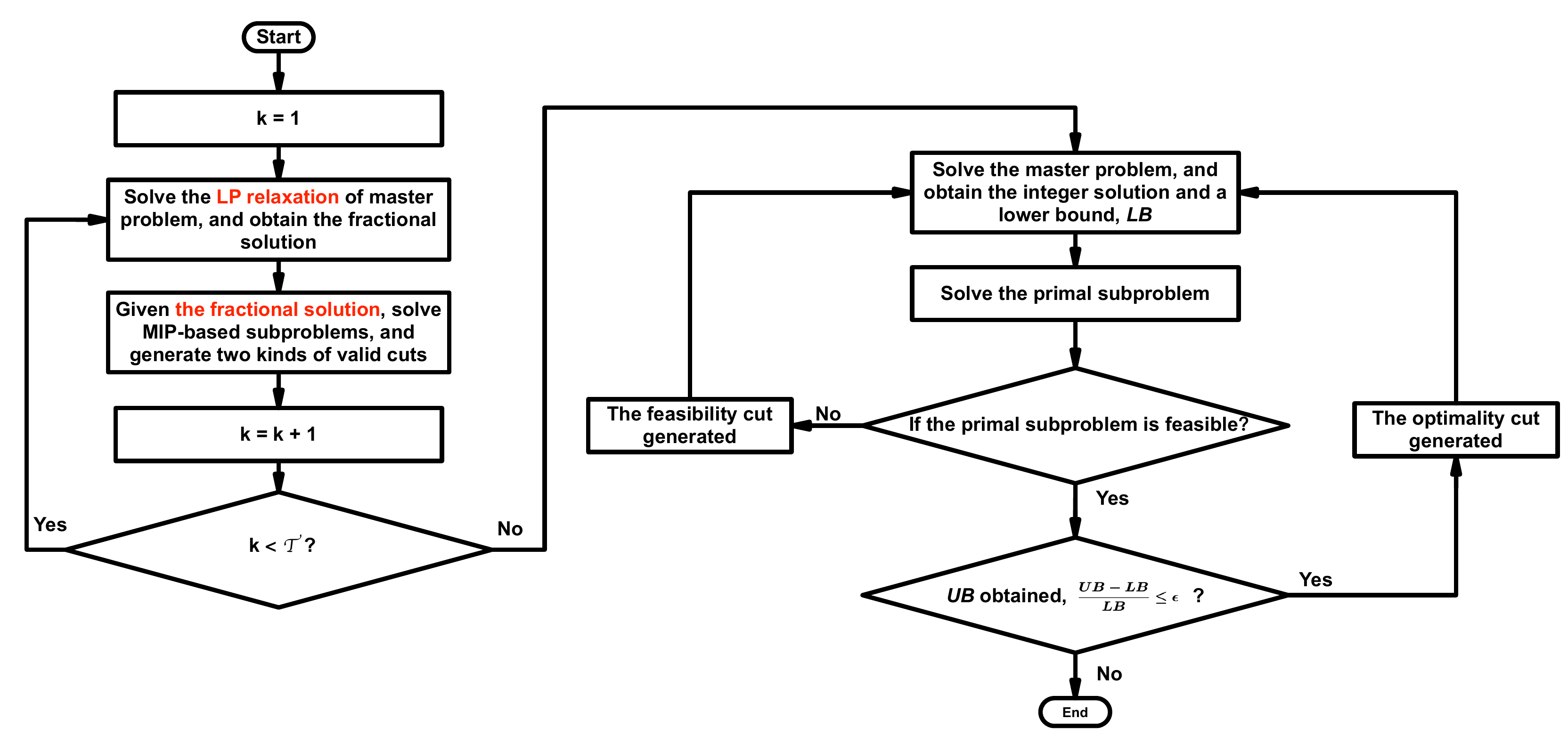}
\caption{\textcolor{black}{The flowchart of the proposed algorithm}}
\label{Solution:algori}
\end{figure*}

\par Thanks to the power of stronger optimality and feasibility cut, compared with the GBD method, the proposed BDD method holds a faster convergence rate for the entire solution process.  Finally, the solution procedures of the proposed BDD is shown in Fig.~\ref{Solution:algori}. Within the first $\mathcal{T}$ iteration, the valid cuts from the proposed BDD method are generated. Besides, the implementation details of GBD are shown on the right hand side of Fig.~\ref{Solution:algori}. \textcolor{black}{ The pseudocode of the proposed algorithm is showcased at Algorithm~\ref{pseudocode:BDD}  for clear illustration. }

\floatname{algorithm}{\color{black}Algorithm}
\begin{algorithm}[h]
 \color{black}
 \caption{\color{black}The Proposed Algorithm}
 \label{pseudocode:BDD}
 \begin{algorithmic}[1]
 \renewcommand{\algorithmicrequire}{\textbf{Input:}}
 \renewcommand{\algorithmicensure}{\textbf{Output:}}
 \STATE  \textbf{\textit{Initialization:} } The stop tolerance $\epsilon$, $k=1$ 
  \FOR{$k\in\mathcal{T}$}
  \STATE Obtain the $x_{ij}^{k\star}, \omega_{j}^{m\star}$ from solving the linear relaxation problem of~\ref{MP}
  \STATE Obtain the dual variables $\lambda_{\omega}^{jm}, \lambda_{x}^{ijk}$ from solving \ref{PSP}
  \IF{The subproblem \ref{PSP} is \textbf{feasible}}
    \STATE Solve the MIP-based subproblem \ref{Problem-LDSP} and generate the strengthened valid optimality cut~\eqref{BDD-Optimality-cut}
  \ELSIF{The subproblem of \ref{MP} is \textbf{infeasible}}
    \STATE Solve the MIP-based feasibility subproblem \ref{FSP} and generate the strengthened valid optimality cut~\eqref{BDD-feasibility-cut}
  \ENDIF
     \STATE {Update $k = k+1$}
  \ENDFOR
  \WHILE{ $(UB-LB)/LB\geq \epsilon$ }
      \STATE Solve the master problem~\ref{MP} and update the \textit{LB}
      \STATE Solve the subproblem~\ref{PSP}
      \IF{The subproblem \ref{PSP} is \textbf{feasible}}
        \STATE Generate the valid optimality cut, and update \textit{UB}
      \ELSIF{The subproblem of \ref{MP} is \textbf{infeasible}}
        \STATE Generate the valid feasibility cut
      \ENDIF
     \STATE {Update $k = k+1$}
     \STATE Go to line 13
  \ENDWHILE
  \RETURN \textit{UB}, \textit{LB}, $k$

 \end{algorithmic} 
 \end{algorithm}

\section{Numerical results}

In this section, extensive simulations are carried out to validate \textit{a)} the superiority of the proposed bi-level model in saving the delivery cost for customers and reducing  the operation cost for the operator,  as well as  \textit{b)} the performance of proposed BDD method in computation time savings and the reduction of the number of iterations compared with the GBD method.

\subsection{Parameter Settings}
To evaluate the performance of proposed algorithm, we test our model on the map of Belgium\footnote{\url{http://www.vrp-rep.org/datasets/item/2017-0001.html}} consisting of the geographic information of 1000 nodes, which is modified from the data set of Yao \cite{yao2021joint}. In terms of parameter settings, revenue from serving a customer is set as $\$9.05$\footnote{\url{https://www.fedex.com/en-us/shipping/one-rate.html}} and the usage cost for vehicle is set as  $\$99$\footnote{\url{https://www.kayak.com/United-States-Car-Rentals.253.crc.html}}.  The following numerical experiments are all implemented 50 times for each instance\footnote{Only considering 5 vehicles. The instances with more vehicles are evaluated in later subsection.}, and statistical results are shown and analyzed. For instance, when the number of customers is 101, we choose these 101 customers randomly from the 1000-node map at each run of simulation experiments. Thanks to the different road topology of randomly selected nodes, the performance of the proposed BDD method over various road topology is demonstrated.  Besides, for brevity, we model the inconvenience function as a two-segment convex piecewise linear function in simulations. $\mathcal{I}(\delta_j)= \max\limits_{n\in\{1,2\}} \gamma_n \delta_j  + \chi_n. $ Note that the inconvenience function can be easily extended to any form of convex optimization problem as explained in above theoretical analysis.  In addition, Table.\ref{dataset} showcases the details of numerical data. We implement all optimization methods with Python 3.7 on Intel Core i9-10980XE CPU 3.00GHz $\times$ 36 with 64 GB of memory.

\begin{table}[h]
            \centering
        \caption{Numerical data for simulations}
    \begin{tabular}{ccccc}
\hline\hline
Parameter  &  Variable  & Value   \\
 \hline\hline
    The slope of the inconvenience function  &    $\gamma_{1,2}$ & 0.5,-0.5    \\  
The intercept of the inconvenience function    &   $\chi_{1,2}$  & -0.01, 0.01\\
 Time flexibility     &  $\Bar{\delta}$  & 1   hour  \\  
 Vehicle usage cost & $c_{v}$ &  99  $\$$  \\
 Delivery fee & $\mathscr{M}_i $ &  9.05  $\$$ \\
 \hline\hline
         \end{tabular}
    \label{dataset}
\end{table}



\subsection{Performance of Bi-level Model}

Compared with the~\ref{VRPTW} model in Appendix.\ref{WO_TF}, the performance of \textbf{BVRP} is evaluated in terms of the operation cost reduction for fleet operator, and the delivery fee saving of customers.

\subsubsection{The superiority of bi-level model in the operation cost reduction of fleet operator}    
\begin{figure}[h]
\centering
 \includegraphics[width=.9\linewidth]{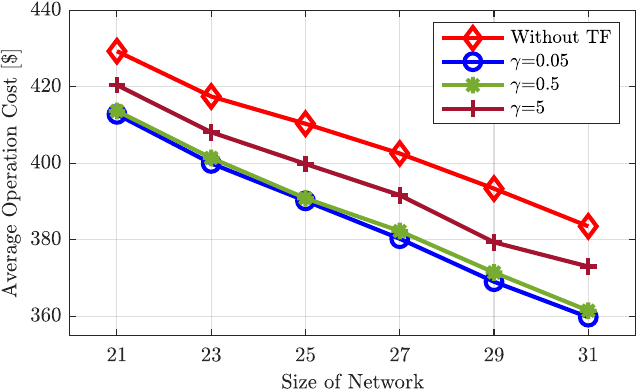}
\caption{The average operation cost reduction over 50 times of experiments }
\label{Performance-TF}
\end{figure}

As illustrated in Fig.~\ref{Performance-TF}, comparison between the proposed \textbf{BVRP} and \ref{VRPTW} \footnote{The details concerning vehicle routing problem without incorporating time flexibility are represented in Appendix \ref{WO_TF}.} showcases that the proposed bi-level optimization model has a good performance on the operation cost saving of fleet operator.  Note that with the increase of the number of customers, the reduction of operation cost  becomes lager, which can explained as that more customers potentially provides more time flexibility for the scheduling of vehicle routing problem. 

\par Furthermore, to test the impact of the different time sensitivity of customers on the reduction of operation cost, three sets of experiments with different value of $\gamma = \{0.05, 0.5,5 \}$ are carried out. Specifically, as illustrated in Fig.~\ref{Performance-TF}, the value of average operation cost reductions increases with the increase of $\gamma$. Since if the fleet operator wants to obtain time flexibility from these customers whose delivery time delay will cause more inconvenience for them, the fleet operator needs to provide higher delivery fee discount. The same reason applies to the simulation results of delivery fee savings in the following subsection. \textcolor{black}{Besides, as the increasing number of network size, more customers are served resulting in the reduction of the entire operational cost.}


\subsubsection{The superiority of bi-level model in the delivery fee savings of customers}  
\begin{figure}[h]
\centering
 \includegraphics[width=.9\linewidth]{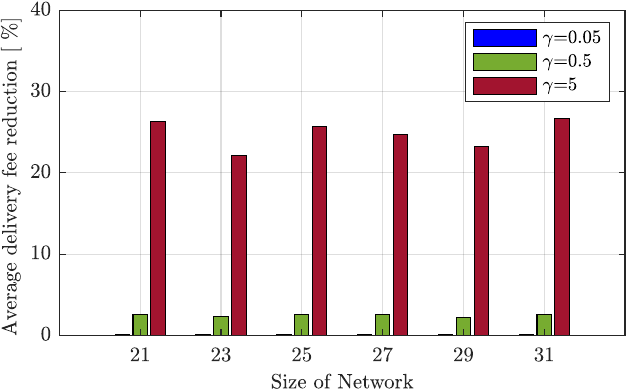}
\caption{The average delivery fee reduction over 50 times of experiments }
\label{Performance-TF-DeliveryFee}
\end{figure}

As shown in Fig.~\ref{Performance-TF-DeliveryFee}, the highest value of average delivery fee reductions of customers has exceeded $25\%$, which demonstrates the power of BVRP in the delivery fee reductions. In addition, the value of average delivery fee reductions becomes larger with the increase of $\gamma$.  The reason has been clarified in above subsection.

\subsection{Performance of Proposed BDD Method}
\begin{figure}[h]
\centering
 \includegraphics[width=.9\linewidth]{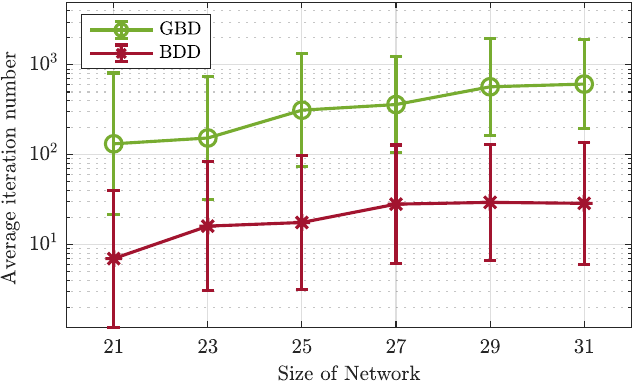}
\caption{\textcolor{black}{The mean and standard deviation value of iteration number of of GBD, and BDD over 50 times of experiments}}
\label{BDD_iteration}
\end{figure}

\textcolor{black}{ In this subsection, the error bar including the information of mean and standard deviation in which the length of error bar is 2 times of the standard deviation, and the specific markers denote the mean values, is employed to show the effectiveness of the proposed BDD.} Compared with the GBD, the performance of proposed BDD method is evaluated in the following two aspects  \textit{(i)} the number of iterations, \textit{(ii)} and the computation time.  Due to the power of stronger Benders cuts, the number of iteration of BDD method will be less than GBD method inherently, which has been illustrated in Fig.~\ref{BDD_iteration}.  

\begin{figure}[h]
\centering
\includegraphics[width=.9\linewidth]{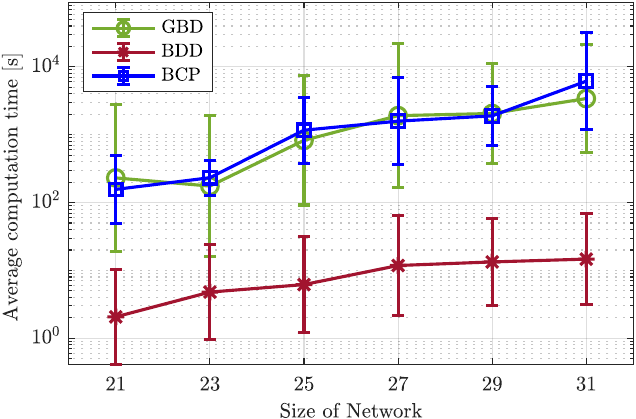}
\caption{\textcolor{black}{The mean and standard deviation value of computation time of GBD, BCP, and BDD over 50 times of experiments}}
\label{BDD_time}
\end{figure}

Consequently, compared with the GBD method, the BDD method, which involves in the computation of mixed integer subproblem, still achieves better performances over all instances in terms of the computation time as illustrated in Fig.~\ref{BDD_time}. \textcolor{black}{In order to further evaluate the performance of the proposed BDD in terms of the reduction of computation time, the branch cut and price method which is the state-of-the-art algorithm in the vehicle routing problem~\cite{desaulniers2016exact,costa2019exact,marques2022branch,mhamedi2022branch,he2019branch,fukasawa2016branch}, is used to verify the effectiveness of the BDD in Fig.~\ref{BDD_time}.  As illustrated in Fig.~\ref{BDD_time}, compared with the state-of-the-art solution method, the proposed BDD method achieves a good performance in reducing the computation time with a decrease of nearly two orders of magnitude.    }


\begin{figure}[h] 
  \centering
  \begin{subfigure}[h]{0.49\textwidth}
    \includegraphics[width=\textwidth]{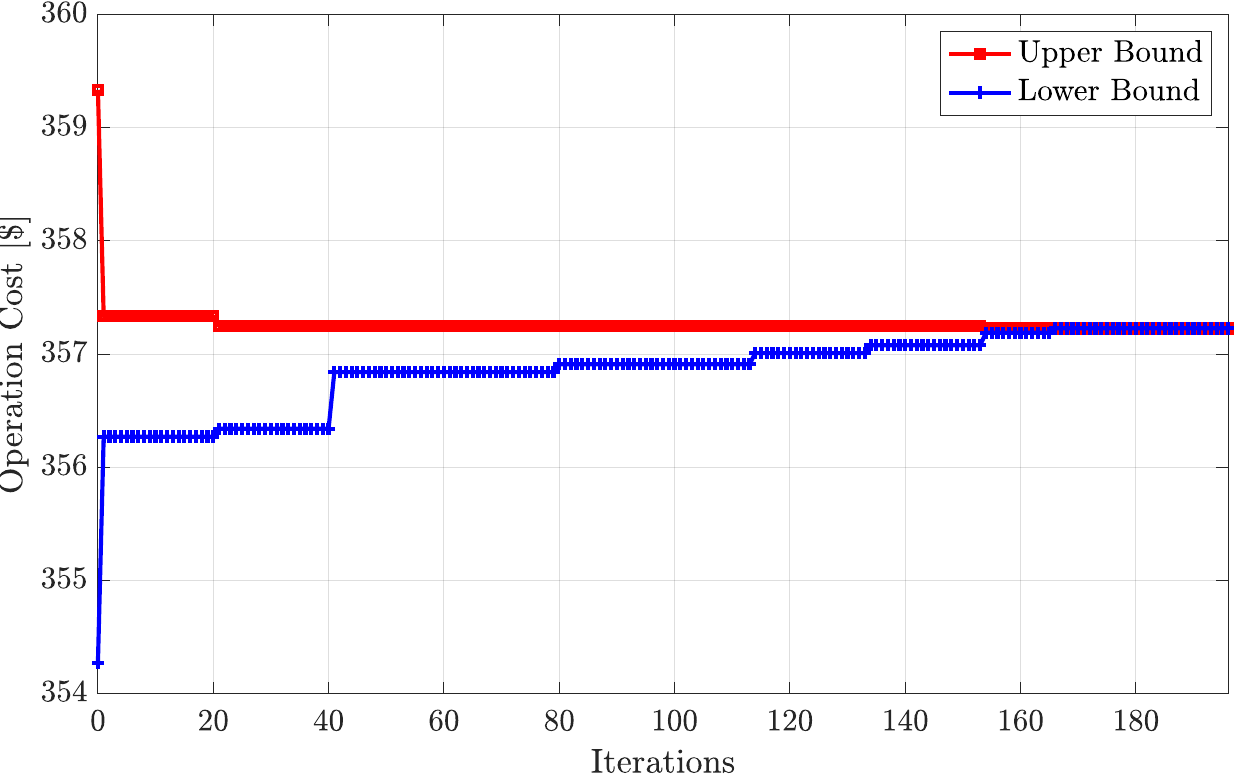} 
    \caption{\color{black}The iteration process of GBD with 31 nodes and 5 vehicles} 
    \label{fig:subfig1} 
  \end{subfigure}
  \hfill 
  \begin{subfigure}[h]{0.49\textwidth}
    \includegraphics[width=\textwidth]{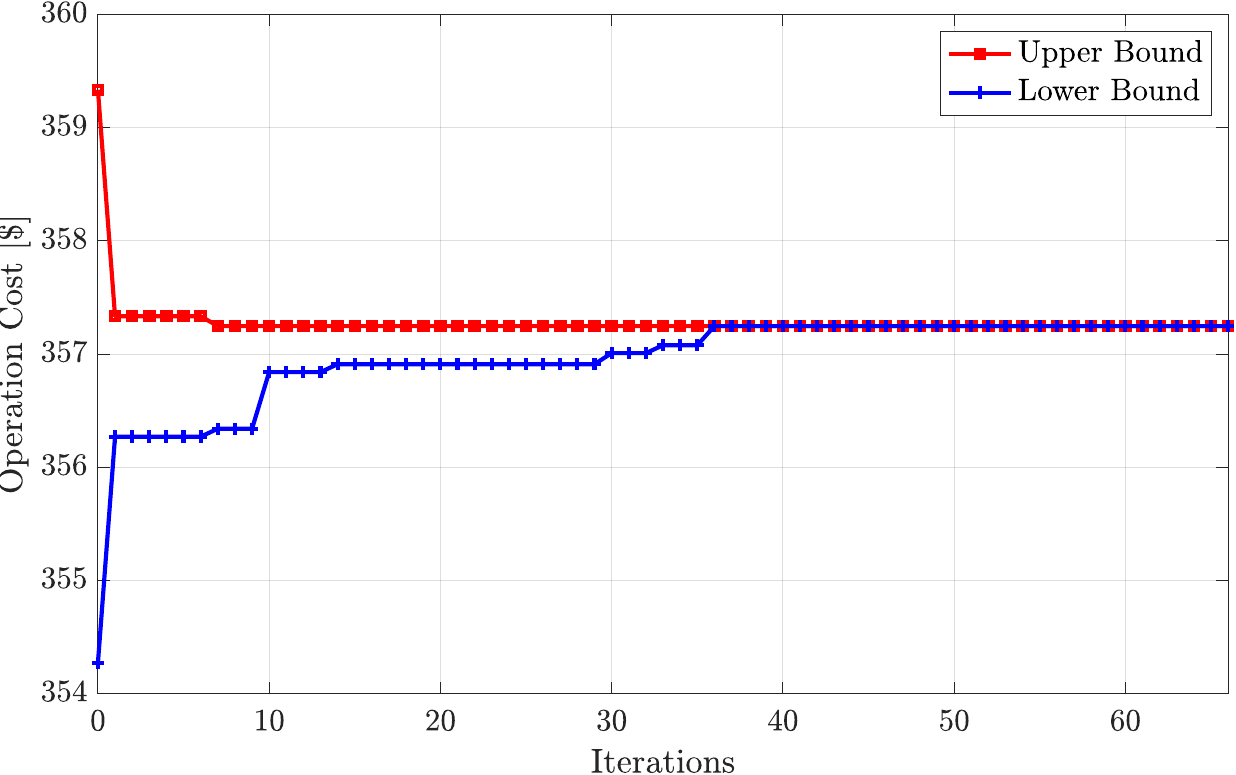} 
    \caption{\color{black}The iteration process of BDD with 31 nodes and 5 vehicles} 
    \label{fig:subfig2} 
  \end{subfigure}
  \caption{\color{black}The iteration process comparison of GBD and BDD with 31 nodes and 5 vehicles. It is worth noting that he upper and lower bounds at iteration 0 are  infinity and negative infinity, respectively.  } 
  \label{fig:iteration} 
\end{figure}

\textcolor{black}{ To clearly demonstrate the performance of BDD, we compare the iteration processes of GBD and BDD in Fig.~\ref{fig:iteration}.  It is evident from the figure that BDD converges to the optimal solution in approximately 35 iterations, whereas GBD requires over 160 iterations to reach the optimal solution.
}

\subsection{Parameter Sensitivity Test on The Time Flexibility }

\begin{figure}[h]
\centering
 \includegraphics[width=.9\linewidth]{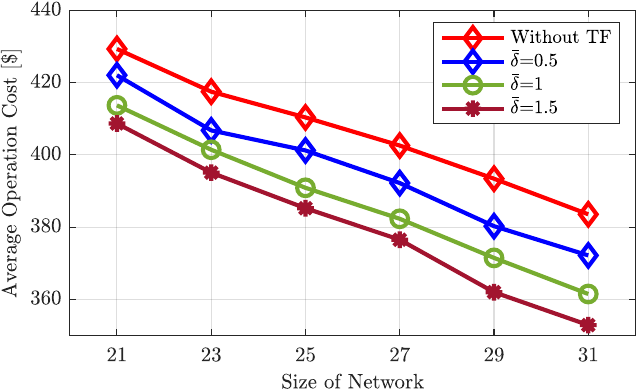}
\caption{Impact of $\bar{\delta}$ on the reduction of operation cost }
\label{Performance-TF-differentWidth-cost}
\end{figure}

\begin{figure}[h]
\centering
 \includegraphics[width=.9\linewidth]{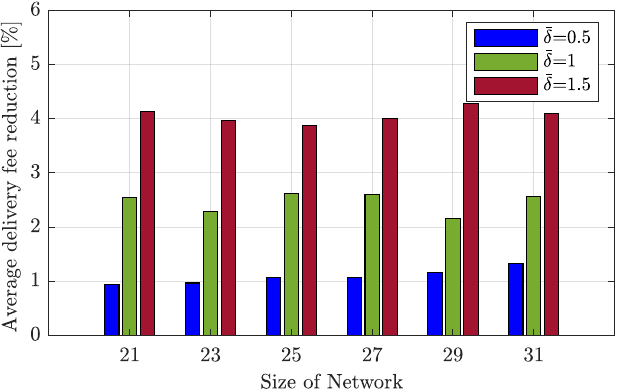}
\caption{Impact of $\bar{\delta}$ on the savings of delivery fee }
\label{Performance-TF-differentWidth-fee}
\end{figure}

To explore the impact of time flexibility on the operation cost reduction, and the delivery fee saving, we conduct 3 sets of simulation experiments with the increasing value of $\bar{\delta}$ from 0.5 hour to 1.5 hour. In Fig.~\ref{Performance-TF-differentWidth-cost}, the operation cost decrease with the increase of $\bar{\delta}$ over all instances, which reveals that more time flexibility is beneficial for the reduction of operation cost. As a result, the larger time flexibility helps the customers save more  delivery fee, which is validated in Fig.~\ref{Performance-TF-differentWidth-fee}.

\subsection{The Scalability of BDD Method}
\begin{figure}[h]
\centering
 \includegraphics[width=.9\linewidth]{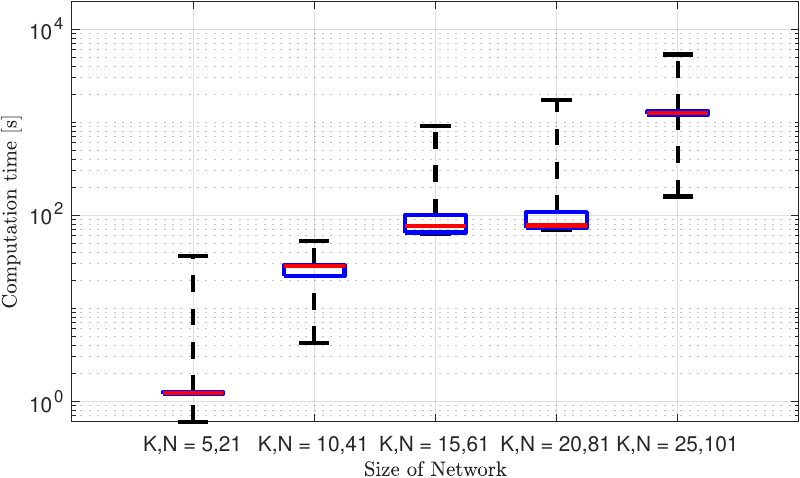}
\caption{The scalability of proposed BDD method}
\label{Performance-TF-scalability}
\end{figure}

In this subsection, extensive experiments are carried out to evaluate the scalability of proposed BDD algorithm by extending the size of instance from 5 vehicles, 21 customers to 25 vehicles, 101 customers. The results in Fig.~\ref{Performance-TF-scalability} demonstrate that the proposed BDD approach can obtain the optimal solution in around 1000 seconds when the size of network is 101 nodes with 25 vehicles considered.

\begin{figure}[h]
\centering
 \includegraphics[width=.9\linewidth]{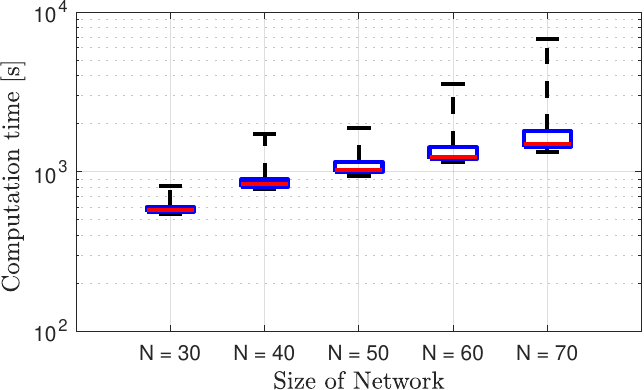}
\caption{\color{black}The performance of the BDD method with the varying number of vehicles}
\label{Performance-varying-number-of-vehicles}
\end{figure}

\textcolor{black}{To assess the performance of the BDD method under the varying number of vehicles, we incrementally introduce vehicles from the set $\{30, 40, 50, 60, 70\}$ in scenarios involving 81 nodes.  The numerical results are shown in Fig.~\ref{Performance-varying-number-of-vehicles}, demonstrating that the BDD method achieves a good performance on large-sized instances.     }


\section{Conclusion}
Considering the delivery time flexibility of customers,  this paper proposes a bi-level optimization framework to characterize the cooperation between the fleet operator and incentive-aware customers for the first time, in which the fleet operator provides price discounts in exchange of customers' time flexibility. Due to the inherent difficulties of bi-level optimization model,  we aim to exactly reformulate the bi-level vehicle routing problem  as a single-level MIP by exploiting the power of KKT optimality condition. In addition, a solution method with low computation complexity is proposed to solve the resulting single-level vehicle routing problem. Finally, to evaluate the performance of the proposed bi-level optimization model and the solution method, we carry out extensive numerical experiments, which validate the performance of proposed bi-level model on the operation cost saving of fleet operator and the delivery fee reduction of customers,  and the efficacy of proposed solution algorithm in computation speed.

\ifCLASSOPTIONcaptionsoff
  \newpage
\fi

\appendices
\section{Proof of Lemma \ref{Lemma-opt-cut}}\label{Proof-opt-cut}

\begin{proof}
	The Benders cut (\ref{BDD-Optimality-cut}) is valid if any $(\Theta, x_{ij}^k, \Omega_{j}^m )$ satisfying 

	\begin{equation}
		\begin{aligned}
			\Theta \geq	\min\limits_{\mathbb{X}_{psp}  } & \Big\{\sum_{k\in \mathcal{K}}\sum_{i\in\mathcal{V}} \sum_{j\in\mathcal{V}}  \eta_{j}  \textit{:} (\ref{Cons_psp}),  \mathcal{X}_{ij}^k,\Omega^m_{j} \in \mathbb{B}\Big\}
		\end{aligned}
	\end{equation}
	also satisfies the cut.	 Given any $(\Theta, x_{ij}^k, \Omega_{j}^m )$ satisfying the above inequality, we have 
		\begingroup\notag
\allowdisplaybreaks
		\begin{align}
			\Theta\geq	 &\min\limits_{\mathbb{X}_{psp}  }  \Big\{\sum_{k\in \mathcal{K}}\sum_{i\in\mathcal{V}} \sum_{j\in\mathcal{V}} \eta_{j}    \textit{:} (\ref{Cons_psp}),  \mathcal{X}_{ij}^k,\Omega^m_{j} \in \mathbb{B}\Big\}\\
			\geq &\max_{\lambda_{\omega}^{jm}, \lambda_x^{ijk} \in  \mathbb{R }  }\Bigg\{ \sum_{j\in \mathcal{R}}\sum_{m\in \mathbb{M}} \lambda_{\omega}^{jm} \omega_j^{m}  +  \sum_{i,j\in \mathcal{V}}\sum_{k\in \mathcal{K}} \lambda_x^{ijk} x_{ij}^{k} \\ &  +    \min\limits_{\mathbb{X}_{psp}  }  \big\{ \sum_{k\in \mathcal{K}}\sum_{i\in\mathcal{V}} \sum_{j\in\mathcal{V}} \eta_{j}   - \sum_{i,j\in \mathcal{V}}\sum_{k\in \mathcal{K}} \lambda_x^{ijk} \mathcal{X}_{ij}^{k} \\&-\sum_{j\in \mathcal{R}}\sum_{m\in \mathbb{M}} \lambda_{\omega}^{jm} \Omega_j^{m}      \textit{:} \eqref{Cons-psp-flow} - \eqref{Cosn-biobj-linearize}  ,  \mathcal{X}_{ij}^k,\Omega^m_{j} \in \mathbb{B} \big\} \Bigg\}\\  
			= &\max_{\lambda_{\omega}^{jm}, \lambda_x^{ijk} \in  \mathbb{R }  }\Bigg\{ \sum_{j\in \mathcal{R}}\sum_{m\in \mathbb{M}} \lambda_{\omega}^{jm} (\omega_j^{m} - \omega_j^{m*})  + \\ & \sum_{i,j\in \mathcal{V}}\sum_{k\in \mathcal{K}} \lambda_x^{ijk} (x_{ij}^{k}-x_{ij}^{k*})     +   \min\limits_{\mathbb{X}_{psp}  }  \big\{ \sum_{k\in \mathcal{K}}\sum_{i\in\mathcal{V}} \sum_{j\in\mathcal{V}}  \eta_{j}  \\ & - \sum_{i,j\in \mathcal{V}}\sum_{k\in \mathcal{K}} \lambda_x^{ijk}  (\mathcal{X}_{ij}^{k} -  x_{ij}^{k*} )    -\sum_{j\in \mathcal{R}}\sum_{m\in \mathbb{M}} \lambda_{\omega}^{jm}  (\Omega_j^{m} \\&-\omega_j^{m*}  )   \textit{:}  \eqref{Cons-psp-flow} - \eqref{Cosn-biobj-linearize}  , \mathcal{X}_{ij}^k,\Omega^m_{j} \in \mathbb{B} \big\} \Bigg\}\\ 
			= &\max_{\lambda_{\omega}^{jm}, \lambda_x^{ijk} \in  \mathbb{R }  }\Bigg\{ \sum_{j\in \mathcal{R}}\sum_{m\in \mathbb{M}} \lambda_{\omega}^{jm} (\omega_j^{m} - \omega_j^{m*})\\&  +  \sum_{i,j\in \mathcal{V}}\sum_{k\in \mathcal{K}} \lambda_x^{ijk} (x_{ij}^{k}-x_{ij}^{k*})    +      \sum_{k\in \mathcal{K}}\sum_{i\in\mathcal{V}} \sum_{j\in\mathcal{V}}   \bar{\eta}_{j} \\ & - \sum_{i,j\in \mathcal{V}}\sum_{k\in \mathcal{K}} \lambda_x^{ijk}  (  \bar{\mathcal{X}}_{ij}^{k} -  x_{ij}^{k*} )  -\sum_{j\in \mathcal{R}}\sum_{m\in \mathbb{M}} \lambda_{\omega}^{jm}  (\bar{\Omega}_j^{m} -\omega_j^{m*}  )          \Bigg\}\\ 
			= &\max_{\lambda_{\omega}^{jm}, \lambda_x^{ijk} \in  \mathbb{R }  }\Bigg\{    \sum_{k\in \mathcal{K}}\sum_{i\in\mathcal{V}} \sum_{j\in\mathcal{V}}  \bar{\eta}_{j}  + \sum_{i,j\in \mathcal{V}}\sum_{k\in \mathcal{K}} \lambda_x^{ijk}  (   x_{ij}^{k} - \bar{\mathcal{X}}_{ij}^{k}  )  \\& +\sum_{j\in \mathcal{R}}\sum_{m\in \mathbb{M}} \lambda_{\omega}^{jm}  ( \omega_j^{m} - \bar{\Omega}_j^{m}   )          \Bigg\}\\   	\geq  & \sum_{k\in \mathcal{K}}\sum_{i,j\in\mathcal{V}}   \bar{\eta}_{j}  + \sum_{i,j\in \mathcal{V}}\sum_{k\in \mathcal{K}} \lambda_x^{ijk}  (   x_{ij}^{k} - \bar{\mathcal{X}}_{ij}^{k}  )  \\&+\sum_{j\in \mathcal{R}}\sum_{m\in \mathbb{M}} \lambda_{\omega}^{jm}  ( \omega_j^{m} - \bar{\Omega}_j^{m}   )  
		\end{align}
	\endgroup	
	where the second line follows from weak duality and the fourth line follows from the optimality of ($ \bar{\eta}_j, \bar{\mathcal{X}}_{ij}^k, \bar{\Omega}_{j}^m $ ). Thus, the proposed Benders cut is valid.

\end{proof}

\section{Proof of Theorem \ref{TightCut}}\label{Proof-Th1}
\begin{proof}
	Given feasible linearization relaxation solution of master problem $\omega_{j}^{m*}, x_{ij}^{k*}$, and dual variables $\hat{\lambda}_{\omega}^{jm}, \hat{\lambda}^{ijk}_{x}$ from \ref{PSP}, we have 
  		\begingroup\notag
\allowdisplaybreaks
		\begin{align}
			\Theta  \geq	&\max_{\lambda_{\omega}^{jm}, \lambda_x^{ijk} \in  \mathbb{R }  }\Bigg\{ \sum_{j\in \mathcal{R}}\sum_{m\in \mathbb{M}} \lambda_{\omega}^{jm} \omega_j^{m*}  +  \sum_{i,j\in \mathcal{V}}\sum_{k\in \mathcal{K}} \lambda_x^{ijk} x_{ij}^{k*} \\ &  +    \min\limits_{\mathbb{X}_{psp}  }  \big\{ \sum_{k\in \mathcal{K}}\sum_{i\in\mathcal{V}} \sum_{j\in\mathcal{V}}  \eta_{j}   - \sum_{i,j\in \mathcal{V}}\sum_{k\in \mathcal{K}} \lambda_x^{ijk} \mathcal{X}_{ij}^{k} \\&-\sum_{j\in \mathcal{R}}\sum_{m\in \mathbb{M}} \lambda_{\omega}^{jm} \Omega_j^{m}      \textit{:} \eqref{Cons-psp-flow} - \eqref{Cosn-biobj-linearize}  ,  \mathcal{X}_{ij}^k,\Omega^m_{j} \in \mathbb{B} \big\} \Bigg\}\\
			\geq	& \sum_{j\in \mathcal{R}}\sum_{m\in \mathbb{M}}  \hat{\lambda}_{\omega}^{jm} \omega_j^{m*}  +  \sum_{i,j\in \mathcal{V}}\sum_{k\in \mathcal{K}} \hat{\lambda}_x^{ijk} x_{ij}^{k*} \\ &  +    \min\limits_{\mathbb{X}_{psp}  }  \big\{ \sum_{k\in \mathcal{K}}\sum_{i\in\mathcal{V}} \sum_{j\in\mathcal{V}}  \eta_{j}   - \sum_{i,j\in \mathcal{V}}\sum_{k\in \mathcal{K}} \hat{\lambda}_x^{ijk} \mathcal{X}_{ij}^{k} \\&-\sum_{j\in \mathcal{R}}\sum_{m\in \mathbb{M}} \hat{\lambda}_{\omega}^{jm} \Omega_j^{m}      \textit{:} \eqref{Cons-psp-flow} - \eqref{Cosn-biobj-linearize}  ,  \mathcal{X}_{ij}^k,\Omega^m_{j} \in \mathbb{B} \big\} \\
				\geq	& \sum_{j\in \mathcal{R}}\sum_{m\in \mathbb{M}}  \hat{\lambda}_{\omega}^{jm} \omega_j^{m*}  +  \sum_{i,j\in \mathcal{V}}\sum_{k\in \mathcal{K}} \hat{\lambda}_x^{ijk} x_{ij}^{k*}   +    \min\limits_{\mathbb{X}_{psp}  }  \Big\{ \sum_{k\in \mathcal{K}}\sum_{i\in\mathcal{V}}\\ & \sum_{j\in\mathcal{V}}  \eta_{j}   - \sum_{i,j\in \mathcal{V}}\sum_{k\in \mathcal{K}} \hat{\lambda}_x^{ijk} \mathcal{X}_{ij}^{k} -\sum_{j\in \mathcal{R}}\sum_{m\in \mathbb{M}} \hat{\lambda}_{\omega}^{jm} \Omega_j^{m}      \textit{:}\\ & \eqref{Cons-psp-flow} - \eqref{Cosn-biobj-linearize}     \Big\}
		\end{align}
	\endgroup	
	Note that the second inequality and third inequality corresponds to strengthened and classic generalized Benders optimality cut respectively. These two optimality cuts are in parallel due to the same slope of these optimality cuts. We use  
	\begin{equation*}
		\begin{aligned}
		    \Xi =& 	 \min\limits_{\mathbb{X}_{psp}  }  \Big\{ \sum_{k\in \mathcal{K}}\sum_{i\in\mathcal{V}} \sum_{j\in\mathcal{V}}  \eta_{j}   - \sum_{i,j\in \mathcal{V}}\sum_{k\in \mathcal{K}} \hat{\lambda}_x^{ijk} \mathcal{X}_{ij}^{k} \\ &-\sum_{j\in \mathcal{R}}\sum_{m\in \mathbb{M}} \hat{\lambda}_{\omega}^{jm} \Omega_j^{m}      \textit{:}  \eqref{Cons-psp-flow} - \eqref{Cosn-biobj-linearize} ,\mathcal{X}_{ij}^k,\Omega^m_{j} \in \mathbb{B}  \Big\}  \\ & -  \min\limits_{\mathbb{X}_{psp}  }  \Big\{ \sum_{k\in \mathcal{K}}\sum_{i\in\mathcal{V}}  \sum_{j\in\mathcal{V}}  \eta_{j}   - \sum_{i,j\in \mathcal{V}}\sum_{k\in \mathcal{K}} \hat{\lambda}_x^{ijk} \mathcal{X}_{ij}^{k}\\ & -\sum_{j\in \mathcal{R}}\sum_{m\in \mathbb{M}} \hat{\lambda}_{\omega}^{jm} \Omega_j^{m}      \textit{:} \eqref{Cons-psp-flow} - \eqref{Cosn-biobj-linearize}   \Big\} 
		\end{aligned}
	\end{equation*}
	  to quantify the tightness between these two optimality cuts.  Due to the positive value of $\Xi$, we conclude that derived optimality cut is $\Xi$ tighter than classic generalized optimality cut. 
\end{proof}

\section{Feasibility subproblem of Benders dual decomposition method}\label{Feasibility-SP}
We denote the decision variables set of feasibility subproblem by  $\mathbb{X}_{fsp}=\{ S\in\mathcal{R}_{+}, \mathbb{X}\setminus \{x_{ij}^{k}, \omega_j^{m} \}, \mathcal{X}_{ij}^k, \Omega_{j}^{m}  \} $.

 \begin{equation*}\tag{\textbf{FSP}}\label{FSP}
	\begin{aligned}
	\min\limits_{  \mathbb{X}_{fsp} } &\sum_{k\in \mathcal{K}}\sum_{i\in\mathcal{V}} \sum_{j\in\mathcal{V}}  S_{1ijk}  +  \sum_{k\in \mathcal{K}}\sum_{i\in\mathcal{V}} \sum_{r\in\{15,16\}} S_{rik}  \\&+\sum_{j\in\mathcal{V}} \sum_{r\in\{2,\cdots,14,17\}} S_{rj} +\sum_{i,j\in \mathcal{V}}\sum_{k\in \mathcal{K}} \lambda_x^{ijk} ( \mathcal{X}_{ij}^{k} - x_{ij}^{k*} ) \\&+ \sum_{j\in \mathcal{R}}\sum_{m\in \mathbb{M} } \lambda_{\omega}^{jm} (\Omega_j^{m}-  \omega_j^{m*}) 
	\end{aligned}
\end{equation*}

\begin{subequations}\notag
    \begin{align}
   \textit{s.t.}\quad  & 	\begin{aligned}
	t_{j}  + S_{1ijk} & \geq  T_{ij}+t_{i} -M(1-\mathcal{X}^k_{ij}), \\ & \forall i\in \mathcal{V}\setminus  v_n, j \in \mathcal{V}\setminus  v_1,k\in\mathcal{K}
	\end{aligned} \\
	 & 	\begin{aligned}
	\tau_j \leq  t_{j} + S_{2j}  ,\quad\forall  j \in \mathcal{V}\setminus  v_1
	\end{aligned} \\
		 & 	\begin{aligned}
	t_j \leq  \tau_{j}  + \delta^{*}_{j} + S_{3j} ,\quad\forall  j \in \mathcal{V}\setminus  v_1
	\end{aligned} \\
		& 			 \left.
		\begin{aligned}
			&	 q_j <=\nabla\mathcal{I}(\delta_j) -q_j -\sigma_j +u_j +S_{4j} \\
			&	 \nabla\mathcal{I}(\delta_j) -q_j -\sigma_j +u_j  <=  S_{5j}+q_j
		\end{aligned}
		\right\}, \forall j\in\mathcal{R} 
		\\	& 
		\left.
		\begin{aligned}
			&0\leq\Bar{\delta}_j-\delta_j+ S_{6j}   \\
			&\Bar{\delta}_j-\delta_j \leq M \Omega^{1}_j + S_{7j}  \\			
			&0\leq u_j+ S_{8j}   \\
			& u_j \leq M (1- \Omega^{1}_j)+ S_{9j}  
		\end{aligned}
		\right\},  \forall j\in\mathcal{R}
		\\&
		\left.
		\begin{aligned}
			&0\leq \delta_j + S_{10j}   \\
			& \delta_j \leq M \Omega^{2}_j + S_{11j}\\			
			&0\leq \sigma_j + S_{12j}\\
			& \sigma_j \leq M (1- \Omega^{2}_j)+ S_{13j}			
		\end{aligned}
		\right\},  \forall j\in\mathcal{R}
		\\&\begin{aligned}
		    \eta_{j}+ S_{14j}	&\geq\mathcal{I}_j(\delta_j)+ u_j \Bar{\delta}_j- \phi^*(\delta^*_j)\\& -M(1-\sum_{k\in\mathcal{K}}\sum_{i\in\mathcal{J}} \mathcal{X}^k_{ij}),\quad\forall j\in\mathcal{R}
		\end{aligned}\\
		& 		\left.
		\begin{aligned}
			& \sum_{j\in\mathcal{V}}   \mathcal{X}^k_{i j}-\sum_{j\in\mathcal{V}}  \mathcal{X}^k_{j i} \leq b_{i} + S_{15ik} \\
			& b_{i} \leq  \sum_{j\in\mathcal{V}}   \mathcal{X}^k_{i j}-\sum_{j\in\mathcal{V}}  \mathcal{X}^k_{j i}+ S_{16ik}
		\end{aligned}
		\right\},  \forall  i \in \mathcal{V}; k\in\mathcal{K} \label{CONS-FSP-flow} \\
      &	\sum_{k\in\mathcal{K}}\sum_{j\in\mathcal{V}} \mathcal{X}^k_{ij} \leq 1 + S_{17i} ,
        \quad\forall i\in\mathcal{R}\\
		&\mathcal{X}_{ij}^k, \Omega_j^m\in \{0,1\}
    \end{align}
\end{subequations}

\section{The mathematical model without incorporating time flexibility}\label{WO_TF}

 \begin{equation}\tag{\textbf{VRPTW}}\label{VRPTW}
\begin{aligned}
\min\limits_{x_{i j}^k, t_j  } \sum_{k\in \mathcal{K}}\sum_{i\in\mathcal{V}} \sum_{j\in\mathcal{V}}  & ( \Gamma T_{ij}+c_{i}) x^k_{i j}   
  \end{aligned}
\end{equation}

  \begin{equation}\notag
\begin{aligned}
\text{s.t.\qquad}  \sum_{j\in\mathcal{V}} & x^k_{i j}-\sum_{j\in\mathcal{V}} x^k_{j i}=b_{i}, \quad \forall  i \in \mathcal{V}; k\in\mathcal{K}\\ &b_{v_1}=1, b_{v_n}=-1, b_{i|i\neq v_1,v_n}=0,
\end{aligned}
\end{equation}
\begin{equation}\notag
    \sum_{k\in\mathcal{K}}\sum_{j\in\mathcal{V}} x^k_{ij} \leq 1,
    \quad\forall i\in\mathcal{R}
\end{equation}

\begin{equation}\notag
	\begin{aligned}
		t_{j}  \geq  &T_{ij}+t_{i} -M(1-x^k_{ij}), \\ & \forall i\in \mathcal{V}\setminus  v_n, j \in \mathcal{V}\setminus  v_1,k\in\mathcal{K}
	\end{aligned}
\end{equation}

\begin{equation}\notag
	\begin{aligned}
	\tau^{l}_{j} \leq	t_{j} \leq  \tau^{u}_{j} ,  \forall  j \in \mathcal{V}\setminus  v_1
	\end{aligned}
\end{equation}

\bibliographystyle{IEEEtran}  
\bibliography{references}

\begin{IEEEbiography}[{\includegraphics[width=1in,height=1.25in,clip,keepaspectratio]{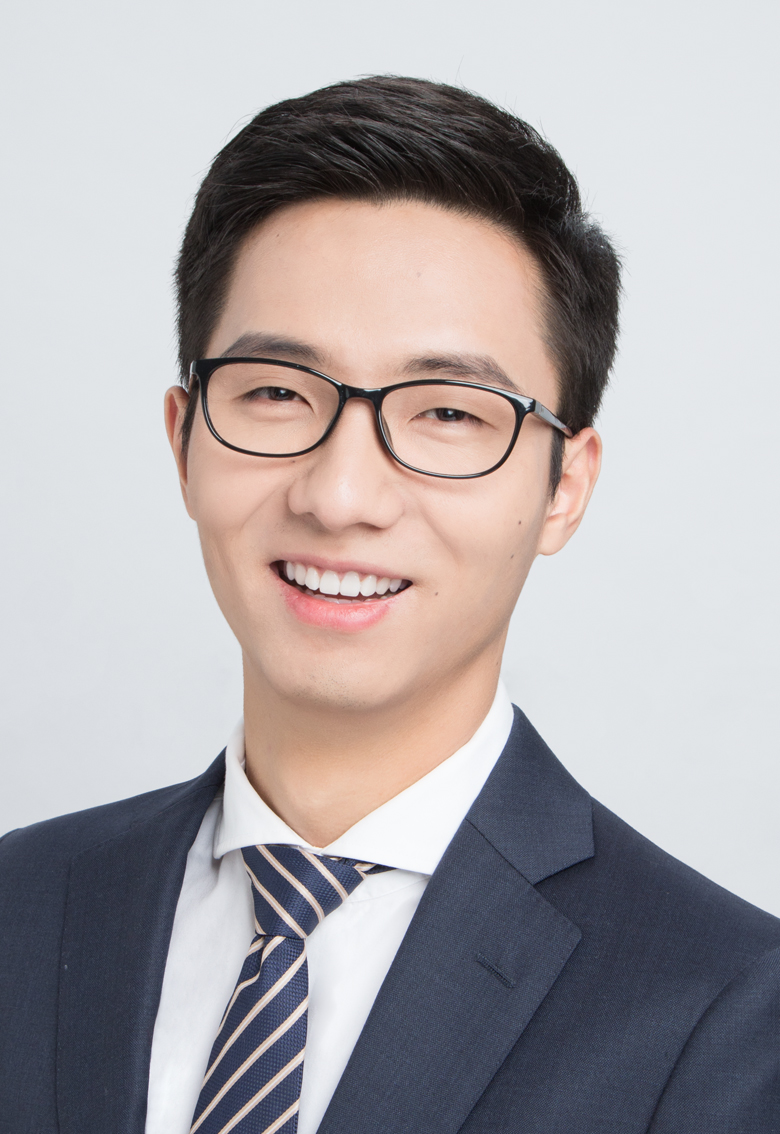}}]{Canqi Yao}
received the B.Eng. degree in electrical engineering from Changsha University of Science and Technology (CSUST), Changsha, Hunan, China, in 2018. He is currently pursuing the Ph.D. degree in mechanical engineering at the Southern University of Science and Technology (SUSTech), Shenzhen, Guangdong, China. He was a visiting PhD student in the Control Systems Technology group, Department of Mechanical Engineering at Eindhoven University of Technology (TU/e), Eindhoven, Netherlands, in 2022. His current research interests include smart grid, electric transportation systems, and optimization theory.
\end{IEEEbiography}

\begin{IEEEbiography}[{\includegraphics[width=1in,height=1.25in,clip,keepaspectratio]{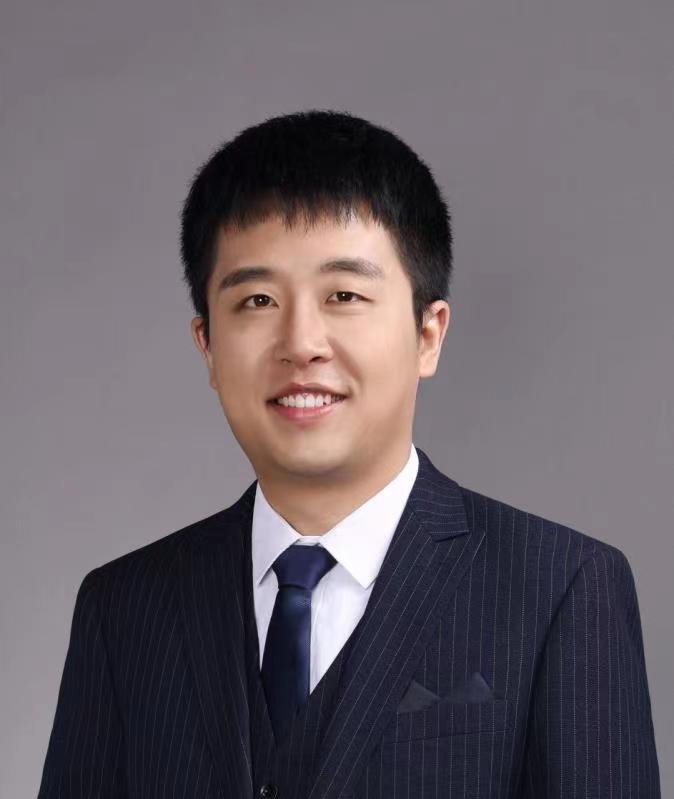}}]{Shibo Chen}
(Member, IEEE) received the B.Eng. degree in electronic engineering from University of Science and Technology of China (USTC), Hefei, China, in 2011 and the Ph.D. degree in electronic and computer engineering from the Hong Kong University of Science and Technology (HKUST), Kowloon, Hong Kong, in 2017. He was a Post- doctoral Fellow with HKUST before joining the Southern University of Science and Technology (SUSTech), Shenzhen, China in 2019 as a Research Assistant Professor. Now he works as a Research
Associate Professor in the School of System Design and Intelligent Manufac- turing, SUSTech. His current research interests include smart grid, intelligent transportation system, optimization and game theory.\end{IEEEbiography}


\begin{IEEEbiography}[{\includegraphics[width=1in,height=1.25in,clip,keepaspectratio]{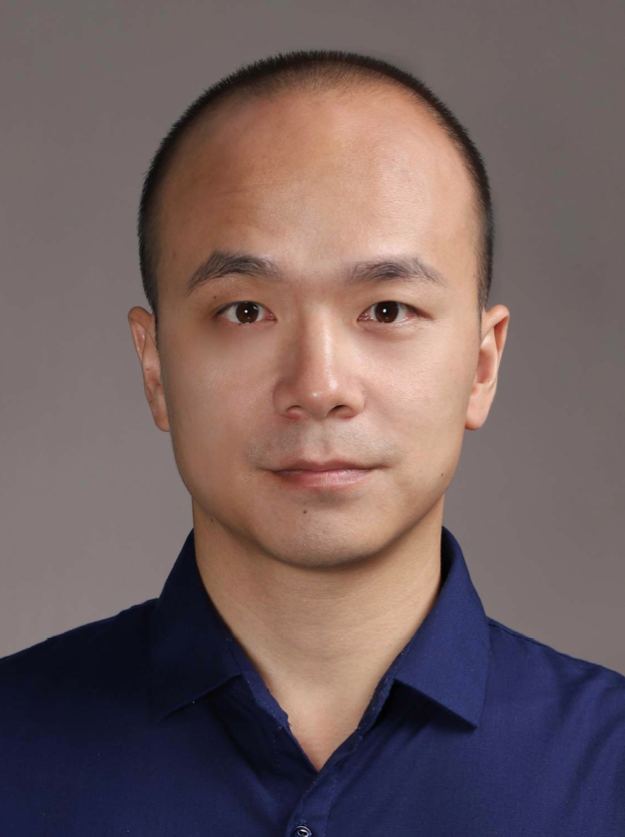}}]{Zaiyue Yang}
 (Senior Member, IEEE) received the B.S. and M.S. degrees from the Department of Automation, University of Science and Technology of China, Hefei, China, in 2001 and 2004, respec- tively, and the Ph.D. degree from the Department of Mechanical Engineering, University of Hong Kong, in 2008. He was a Postdoctoral Fellow and Re- search Associate with the Department of Applied Mathematics, Hong Kong Polytechnic University, before joining the College of Control Science and Engineering, Zhejiang University, Hangzhou, China,
in 2010. Then, he joined the School of System Design and Intelligent Manufacturing, Southern University of Science and Technology, Shenzhen, China, in 2017. He is currently a Professor there. His current research interests include smart grid, signal processing and control theory. Prof. Yang is an associate editor for the IEEE Transactions on Industrial Informatics.\end{IEEEbiography}

\end{document}